\documentclass[a4,14pt]{article}
\usepackage[margin=1in]{geometry}
\usepackage{subcaption}
\usepackage{amsmath}
\usepackage{hyperref}
\usepackage{amssymb}
\usepackage{braket}
\usepackage{upgreek}
\usepackage{graphicx}
\usepackage{float}
\usepackage{cite}
\usepackage{esvect}
\begin{document}
\pagenumbering{arabic}
\title{Mass Varying Neutrino Oscillation in Teleparallel Gravity}
\author{H. Mohseni Sadjadi\footnote{mohsenisad@ut.ac.ir} and H. Yazdani Ahmadabadi\footnote{hossein.yazdani@ut.ac.ir}
	\\ {\small Department of Physics, University of Tehran,}
	\\ {\small P. O. B. 14395-547, Tehran 14399-55961, Iran}}
\maketitle
\begin{abstract}
In the Cartan teleparallel formulation of gravity, a scalar field coupled to the torsion tensor can acquire an environment-dependent
effective mass. This gives rise to a screening effect: inside dense bodies such as the Sun, the scalar field is suppressed,
causing the mass varying neutrino mass to vary with local density. We investigate the impact of this torsion-induced
screening on solar neutrino oscillations. The mechanism modifies the standard MSW resonance condition in a distinct way,
leading to detectable shifts in the flavor conversion probability. Using data from solar neutrino experiments
(Super-Kamiokande, Borexino, SNO), we place constraints on the torsion–scalar coupling parameters.
Our work provides concrete test of teleparallel gravity through mass-varying neutrino oscillations.
 \end{abstract}

\section{Introduction}\label{sec1}
Experimental evidences \cite{SNOEx,Super-Kamiokande-Fukuda} has convincingly demonstrated that neutrinos undergo flavor conversion as
they propagate in a vacuum or even through matter, a phenomenon that can only occur if they possess non-zero masses.
This fascinating behavior indicates the necessity for an extension of the standard model (SM) or a more comprehensive examination of
the various interactions that neutrinos may experience.
Numerous mechanisms beyond the standard model (BSM) have been proposed to account for the non-zero mass of neutrinos, including the
seesaw mechanism \cite{Fardon,Bi}, the introduction of extra dimensions \cite{Bi}, and theories of supersymmetry \cite{SuSy}.

A well-recognized modification of flavor evolution in matter is the MSW effect \cite{Mikh-Smir-MSW,Mikh-Smir,Wolf-MSW,Wolf}, wherein
the forward scattering of neutrinos off electrons generates an effective potential that modifies the probability of flavor conversion.
Additionally, it is essential to explore whether other types of interactions, particularly those mediated by scalar fields, could produce
analogous or competing effects in the calculations of neutrino oscillations
 \cite{Stephenson,Halprin,Kaplan,Barger-DE,Fardon,MohseniSadjadi,MohseniSadjadi1,MohseniSadjadi2,YarAhm,Mazia,Hung,Brookfield,Geng}.

The interaction of scalar fields with neutrinos has been explored within the framework of mass-varying neutrino (MaVaN) models, which are
often inspired by the intriguing relationship between the scales of dark energy and neutrino masses \cite{Stephenson,Halprin,Fardon}.
These models suggest that the masses of neutrinos may evolve dynamically as a result of their coupling with a cosmological scalar field,
 thereby potentially establishing a connection between cosmic acceleration and neutrino physics \cite{Stephenson,Halprin,Kaplan,Barger-DE,Fardon,MohseniSadjadi,MohseniSadjadi1,MohseniSadjadi2,YarAhm,Mazia,Hung,Brookfield,Geng}.
This relationship implies that the variability of neutrino masses could significantly influence cosmological evolution.
In such contexts, the scalar field may exhibit sensitivity to local matter density, leading to environment-dependent variations
 in neutrino masses and altering their oscillation patterns in a manner analogous to the MSW effect \cite{MohseniSadjadi-Yazdani1,MohseniSadjadi-Yazdani2,MohseniSadjadi-Yazdani3,MohseniSadjadi-Yazdani4,MohseniSadjadi-Yazdani5,Sadjadi-Khosravi,Chakraborty}.

Expanding upon this concept, certain scalar-tensor theories of gravity, including the Chameleon \cite{Khoury,Chameleon,Brax} and
Symmetron models \cite{sym,sym-Hinterbichler}, have been investigated.
In these frameworks, scalar fields equipped with screening mechanisms facilitate conformal couplings between neutrinos and the gravitational
sector \cite{MohseniSadjadi-Yazdani1,MohseniSadjadi-Yazdani2,MohseniSadjadi-Yazdani3,MohseniSadjadi-Yazdani4,Sadjadi-Khosravi,Chakraborty}.
The Chameleon and Symmetron models allow for the coupling of the scalar field to matter density, thereby ensuring that the scalar field
adheres to local gravitational constraints while simultaneously exerting an influence on neutrino physics in regions of low density.

However, most of the above MaVaN and screening models share a common feature: the scalar field couples directly to the local matter
density (or the trace of the stress-energy tensor). This leaves open an important theoretical gap: could a screening effect arise
from a purely geometric origin, namely from torsion, rather than from a direct coupling to matter? And if so, how would such a
torsion-induced screening affect neutrino oscillations?

An alternative hypothesis regarding the origin of neutrino mass may stem from the couplings of neutrinos and torsion both with an unknown
particle, specifically a scalar field, within the context of the teleparallel equivalent of general relativity (TEGR)
\cite{Capolupo1,Capolupo2,Panda,MohseniSadjadi-Yazdani6}.
Researches in this area suggests that the interaction of the scalar field with spacetime torsion acts as an effective potential
that affects neutrino propagation, thereby altering their kinematic characteristics.
In contrast to conventional GR, which is based on curvature, TEGR reinterprets gravitational phenomena through the concept of torsion,
employing the torsion scalar $T$ within a Riemann–Cartan spacetime framework characterized by
the Weitzenb$\ddot{\text{o}}$ck connection\cite{Weitzenbock,Teleparallel-2,Teleparallel-1,Teleparallel-3,Teleparallel-4,Teleparallel-5,Teleparallel-6,Teleparallel-7}.

Despite the growing interest in teleparallel gravity, the specific question of how a torsion–scalar coupling induces
a density-dependent screening of neutrino masses and how this effect can be tested using solar neutrino oscillations
has not been addressed in the literature. This is precisely the gap that the present work fills.

Within the present study, the impact of the coupling between torsion and a scalar field on neutrino oscillations is examined,
 with an emphasis on the considerations of teleparallel gravity integrated into the formalism through the \textit{effective neutrino mass term}.
Analytical expressions are derived for both the phase shifts associated with vacuum neutrino oscillations and the mixing parameters inside matter.
Thus, these modifications carry substantial implications for fundamental vacuum oscillations and the traditional MSW
 resonance mechanism, highlighting the significance of torsion-scalar field coupling as an essential factor in astrophysical scenarios.
Furthermore, by analyzing experimental data from solar neutrino observations, we impose constraints on the model parameters, providing
 new insights into the mass-varying flavor conversion of neutrinos within the TEGR framework.
The constraints derived are in agreement with the LMA-MSW solution for the solar neutrino problem, suggesting that the TEGR framework
is capable of fitting observational data while maintaining consistency with conventional oscillation predictions.

The structure of the paper is outlined as follows:
In Section \ref{sec2}, we derive the profile of the scalar field both within and beyond a spherical body, such as the Sun,
 based on the coupling between the scalar field and torsion.
Section \ref{sec3} focuses on the examination of neutrino flavor conversion in the presence of the scalar field.
In Section \ref{sec4}, we provide numerical results and discuss the constraints imposed by solar neutrino experiments.
Section \ref{sec5} presents our findings and engages in a discussion surrounding them.
Finally, concluding remarks are presented in Section \ref{sec6}.

Throughout this paper, we work in natural units where $\hbar = c = 1$.

\section{Coupled Scalar Field in Teleparallel Gravity}\label{sec2}
This work investigates a cosmological framework within the teleparallel formulation of gravity.
The model posits a Universe whose energy content is primarily composed of a dynamical dark energy component, represented by a scalar field $\phi$, coupled with pressureless baryonic matter.
The gravitational and field dynamics are governed by the action
\begin{eqnarray}\label{eqn1}
S=\int d^4 x ~\mathcal{E} \left[\frac{M_p^2 T}{2} + \frac{1}{2} \partial_\mu \phi \partial^\mu \phi + \frac{1}{2} \epsilon T\phi^2 + \kappa_i f(\phi) \bar{\nu}_i \nu_i - i\bar{\nu}_i \gamma^\mu D_\mu \nu_i + \mathcal{L}_m(\Psi^{(i)},e^{\hat{a}}_{\mu})\right],
\end{eqnarray}
where $M_p = 1/\sqrt{8\pi G}$ is the reduced Planck mass, $\mathcal{E}$ is the determinant of the tetrad field $e^{\hat{a}}_\mu$, and $\mathcal{L}_m$ represents the Lagrangian for matter components $\Psi^{(i)}$.
Here, $T$ denotes the torsion scalar that defines the teleparallel gravitational sector.
A key feature of this model is a non-minimal coupling between the scalar field and gravity, introduced through the term proportional to $\epsilon T \phi^2$ , where $\epsilon$ is a coupling constant.
Furthermore, the model includes an interaction between scalar field $\phi$ and neutrino fields $\nu_i$, described by the Yukawa-like coupling $\kappa_i f(\phi) \bar{\nu}_i \nu_i$, where $f(\phi)$ is a coupling function.

Variation of the action with respect to $\phi$ gives
\begin{eqnarray}\label{eqn2}
\Box\phi - \epsilon T \phi - \kappa_i f_{,\phi}(\phi) \bar{\nu}_i \nu_i  = 0.
\end{eqnarray}
Under the assumption that the neutrino number density, $n_\nu = \bar{\nu}_i \nu_i$, is negligible, the last interaction term in the field equation can be justifiably disregarded.
This simplification arises from the consideration that in specific cosmological regimes, the energy density contribution from neutrinos becomes sub-dominant, allowing their direct scalar coupling to be treated as a next-order effect.

To facilitate the resolution of the field equation, the explicit functional form of the torsion scalar $T$ must be determined.
This requires the specification of a spacetime geometry, for which we adopt a static, spherically symmetric metric ansatz \cite{Tamanini}
\begin{eqnarray}\label{eqn3}
ds^2 = a^2(r) dt^2 - b^2(r) dr^2 -r^2 d\theta^2 - r^2 \sin^2\theta d\varphi^2.
\end{eqnarray}
In the framework of teleparallel gravity, the fundamental variable is the tetrad field $e^{\hat{a}}_\mu$, instead of the metric (\ref{eqn3}).
For the specified metric, a corresponding tetrad component $e^{\hat{a}}_\mu$ can be formulated as
\begin{eqnarray}\label{eqn4}
e^{\hat{a}}_\mu =
\begin{pmatrix}
a(r) & 0 & 0 & 0 \\
0 & b(r) \sin\theta \cos\varphi & r \cos\theta \cos\varphi & -r \sin\theta \sin\varphi \\
0 & b(r) \sin\theta \sin\varphi & r \cos\theta \sin\varphi & r \sin\theta \cos\varphi \\
0 & b(r) \cos\theta & -r \sin\theta & 0
\end{pmatrix}
.
\end{eqnarray}
This particular tetrad is not arbitrary; it is meticulously constructed to be aligned with the symmetry of the system.
A significant consideration in the teleparallel gravity is the separation of gravitational from inertial effects.
The selected tetrad is a ``good'' or ``proper'' one in the sense that it inherently minimizes spurious inertial contributions, thereby providing a pure gauge representation of the gravitational field.
This makes it the preferred and physically meaningful choice for analyzing configurations with spherical symmetry \cite{Tamanini}.

The inverse of the tetrad field, denoted by $e^\mu_{\hat{a}}$, is subsequently defined by the orthonormality condition $e^{\hat{a}}_\mu e^\mu_{\hat{b}} = \delta^{\hat{a}}_{\hat{b}}$, which ensures that the tetrad and its inverse map between the spacetime and tangent space frames consistently \cite{Maluf}.
For the specific tetrad under consideration, the components of the inverse field are given by the following matrix representation:
\begin{eqnarray}\label{eqn5}
e^\mu_{\hat{a}} =
\begin{pmatrix}
\frac{1}{a(r)} & 0 & 0 & 0 \\
0 & \frac{\sin\theta \cos\varphi}{b(r)} & \frac{\sin\theta \sin\varphi}{b(r)} & \frac{\cos\theta}{b(r)} \\
0 & \frac{\cos\theta \cos\varphi}{r} & \frac{\cos\theta \sin\varphi}{r} & -\frac{\sin\theta}{r} \\
0 & -\frac{\sin\varphi}{r \sin\theta} & \frac{\cos\varphi}{r \sin\theta} & 0
\end{pmatrix}
.
\end{eqnarray}

The torsion scalar $T$, which quantifies the teleparallel equivalent of curvature, is computed from the specified tetrad field (\ref{eqn4}) and its inverse (\ref{eqn5}) \cite{Tamanini}.
For the given spherically symmetric configuration, the general form of this scalar assumes the following explicit form
\begin{eqnarray}\label{eqn6}
T = \frac{2}{r^2} \left[1 - \frac{1}{b(r)}\right] \left[1 - \frac{1}{b(r)} - \frac{2 r}{a(r) b(r)}  \frac{da(r)}{dr} \right].
\end{eqnarray}
In $f(T)$ gravity, the choice of tetrad is crucial because the theory is sensitive to inertial effects encoded in the spin connection.
The proper tetrad is selected to minimize artificial inertial contributions, making it physically meaningful for studying spherically symmetric spacetimes.
The diagonal tetrad in spherical coordinates introduces inertial forces that contaminate the torsion scalar $T$.
The good tetrad, however, is constructed such that the spin connection $\omega^{\hat{a}}_{\hat{b}\mu}$ cancels these inertial terms, yielding a torsion scalar $T$ that depends only on the gravitational field \cite{Ruggiero}.

\subsection{Scalar Field Behavior}\label{subsec2-1}
The present analysis investigates the scalar field configuration arising from a localized, spherically symmetric source of mass $M_\odot$ and radius $R_\odot$, situated within an environment of dilute density.
As a second-order differential equation (Eq.(\ref{eqn2})), the system necessitates the specification of two boundary conditions to yield a unique physical solution \cite{MohseniSadjadi-Yazdani5}.
The first condition arises from the requirement of regularity at the coordinate origin; to preclude a physical singularity at $r = 0$, we impose that the radial derivative vanishes: $\frac{d\phi}{dr}|_{r=0} \to 0$.
The second condition is applied asymptotically, enforcing that the scalar field approaches its ambient background value far from the source: $\phi \to \phi_\infty$ at $r\to \infty$.
This asymptotic condition is physically well-motivated, as it ensures that the field gradient—and consequently the fifth force mediated by $\phi$—vanishes at spatial infinity.

According to the interior Scwharzschild metric, i.e., for $a(r^\prime) = \frac{1}{2} \left(3\sqrt{1 - r^\prime_s} - \sqrt{1- r^{\prime 2} r^\prime_s}\right)$ and $b(r^\prime) = \left(1- r^{\prime 2} r^\prime_s\right)^{-1/2}$, the relation (\ref{eqn6}) gives
\begin{eqnarray}\label{eqn7}
	T_{\text{in}}(r^\prime) = \frac{2}{r^{\prime 2} R^2_\odot} \left(1 - \sqrt{1 - r^{\prime 2} r^\prime_s}\right) \left(1 - \sqrt{1 - r^{\prime 2} r^\prime_s} - \frac{2 r^{\prime 2} r^\prime_s}{3 \sqrt{1 - r^\prime_s} - \sqrt{1 - r^{\prime 2} r^\prime_s}}\right).
\end{eqnarray}
It is important to note that the radial coordinate $r^\prime$ appearing in this equation is dimensionless, representing the fractional radius normalized by the solar radius, $R_\odot$.
Expanding this relation around the small $r^\prime_s$ leads to $T_{\text{in}}(r^\prime) \simeq -\frac{r^{\prime 2} r_s^{\prime 2}}{2 R_{\odot}^2}  + \frac{r^{\prime2} \left(r^{\prime 2} - 3\right) r_s^{\prime 2}}{4 R_\odot^2} + \mathcal{O}\left(r_s^{\prime 4}\right)$.
For the solar case, we have $\frac{T_{\text{in}}^{(2)}}{T_{\text{in}}^{(1)}} \ll 10^{-6}$.
Based on Eq.(\ref{eqn2}), the explicit form of the field equation for \textbf{inside} the sphere (i.e., $r < R_\odot$) is given by
\begin{eqnarray}\label{eqn8}
\begin{split}
&\left(1 - r_s^\prime r^{\prime 2}\right) \frac{d^2 \hat{\phi}}{dr^{\prime 2}} - \frac{\left[\left(4 r^{\prime 4} - 27 r^{\prime 2}\right) r_s^{\prime 2} + 3 r_s \left(r^{\prime 2} \left(7 - \sqrt{1 - r_s^\prime} \sqrt{1 - r^{\prime 2} r_s^\prime}\right) + 6 \right) - 16\right]}{r^{\prime 3} r_s^\prime + r^\prime (8 - 9 r_s^\prime)} \frac{d\hat{\phi}}{dr^\prime} + \\& \epsilon R_\odot^2 T_{\text{in}}(r^\prime) \hat{\phi} = 0,
\end{split}
\end{eqnarray}
where we have redefined dimensionless parameters
\begin{eqnarray}\label{eqn9}
\hat{\phi}(r) = \frac{\phi(r)}{\phi_{\text{in}}(0)} ,~~~~~~~~ r^\prime = \frac{r}{R_\odot} ,~~~~~~~~  r_s^\prime = \frac{r_s}{R_\odot}.
\end{eqnarray}
It is worth noting that the normalization (\ref{eqn9}) for the scalar field seems vital because the present model of TEGR does not fix the field's value at the center of the sphere (i.e., at $r' = 0$), leaving $\phi_{\text{in}}(0)$ as a parameter that can be absorbed into others.
Consequently, the dimensionless field $\hat{\phi}(r')$ represents the scalar field behavior relative to this central value $\phi_{\text{in}}(0)$.
The large values of $\hat{\phi}(r^\prime)$ outside the Sun do not necessarily imply large physical values of the scalar field $\phi(r)$ itself; they indicate that the field profile grows relative to its central boundary value.

Although normalizing with respect to the reduced Planck mass $M_p$ is mathematically possible, it would leave the normalized field $\hat{\phi}(r)$ unchanged, since it is defined as the ratio of $\phi(r)$ to $\phi_{\text{in}}(0)$—a quantity invariant under rescaling by any fundamental constant.
Nevertheless, the current normalization is advantageous, as it intrinsically accounts for the theory’s undetermined boundary value $\phi_{\text{in}}(0)$, offering a more natural and intuitive depiction of the field’s spatial variation in relation to its central value.

The profile of the scalar field \textbf{inside} the medium (i.e., $r < R_\odot$) is therefore given by
\begin{eqnarray}\label{eqn10}
\hat{\phi}_{\text{in}}(r^\prime) = e^{-\frac{1}{8} r^\prime_s r^{\prime 2} \left(\sqrt{8 \epsilon + 25} - 5\right)} \, _1F_1\left(\frac{3}{4} \left(1-\frac{5}{\sqrt{8 \epsilon +25}}\right);\frac{3}{2};\frac{1}{4} r^\prime_s r^{\prime 2} \sqrt{8 \epsilon + 25}\right),
\end{eqnarray}
where $_1F_1\left(a;b;z\right)$ is Kummer confluent hypergeometric function.

For the exterior Scwharzschild metric, however, we have $a(r^\prime) = b^{-1}(r^\prime) = \left(1 - \frac{r^\prime_s}{r^\prime}\right)^{1/2}$, so, the good tetrad (\ref{eqn4}) gives
\begin{eqnarray}\label{eqn11}
T_{\text{out}}(r^\prime) =- \frac{2}{r^{\prime 2} R_\odot^2} \frac{\left[1 - \sqrt{1 - \frac{r^\prime_s}{r^\prime}} \right]^2}{\sqrt{1 - \frac{r^\prime_s}{r^\prime}}},
\end{eqnarray}
which at large distances, i.e, for $r^\prime_s \ll r^\prime$ (weak-field limit), is approximated by $T_{\text{out}}(r^\prime) \simeq -\frac{r_s^{\prime 2}}{2 R^2_\odot r^{\prime 4}}  -\frac{r_s^{\prime 3}}{2 R^2_\odot r^{\prime 5}} + \mathcal{O}\left(r_s^{\prime4}\right)$.
For the solar case, we have $\frac{T_{\text{out}}^{(2)}}{T_{\text{out}}^{(1)}} \ll 10^{-6}$.
Based on Eq.(\ref{eqn2}), the explicit form of the field equation for \textbf{outside} the sphere (i.e., $r > R_\odot$) is given by
\begin{eqnarray}\label{eqn12}
\left(1 - \frac{r_s^\prime}{r^\prime}\right) \frac{d^2\hat{\phi}}{dr^{\prime 2}} + \frac{1}{r^\prime} \left(2 - \frac{r_s^\prime}{r^\prime}\right) \frac{d\hat{\phi}}{dr^\prime} + \epsilon R^2_\odot T_{\text{out}}(r^\prime) \hat{\phi} =0 .
\end{eqnarray}

By applying continuity of the field and its derivative at the boundary of the spherical source ($r^\prime=1$), the exterior solution for the scalar field is approximately determined to be
\begin{eqnarray}\label{eqn13}
\begin{split}
\hat{\phi}_{\text{out}} (r^\prime) & = \frac{\exp \left[-\frac{r_s^\prime \left(r^\prime \left(4 \sqrt{2 \epsilon + 1} + \sqrt{8 \epsilon + 25} - 9\right) + 4 \left(\sqrt{2 \epsilon + 1} + 1 \right)\right)}{8 r^\prime}\right]}{4 \sqrt{2 \epsilon + 1}} \times \\& \biggr[\, _1F_1\left(\frac{7}{4} - \frac{15}{4 \sqrt{8 \epsilon + 25}};\frac{5}{2};\frac{1}{4} r_s^\prime \sqrt{8 \epsilon + 25}\right) \left(\sqrt{8 \epsilon + 25} - 5\right) \left(e^{r_s^\prime \sqrt{2 \epsilon + 1}} - e^{\frac{r_s^\prime \sqrt{2 \epsilon + 1}}{r^\prime}}\right) \\& + \, _1F_1\left(\frac{3}{4} \left(1 - \frac{5}{\sqrt{8 \epsilon + 25}}\right);\frac{3}{2};\frac{1}{4} r_s^\prime \sqrt{8 \epsilon + 25}\right) \bigg(\left(2 \sqrt{2 \epsilon + 1} + \sqrt{8 \epsilon + 25} - 3\right) e^{\frac{r_s^\prime \sqrt{2 \epsilon + 1}}{r^\prime}}  \\& +  \left(2 \sqrt{2 \epsilon + 1} - \sqrt{8 \epsilon + 25} + 3\right) e^{r_s^\prime \sqrt{2 \epsilon + 1}}\bigg)\biggr].
\end{split}
\end{eqnarray}
To investigate the phenomenological consequences for gravitational tests, it is necessary to determine the configuration of this scalar field in the vicinity of astrophysical bodies, with the Sun serving as a primary test case.
For this specific application, the analysis can be significantly simplified by employing the Newtonian approximation.
Within this regime, it is consistent to neglect both non-linear gravitational effects and the back-reaction of the scalar field on the spacetime metric, as their contributions are sub-dominant for the systems under consideration.

To ensure the scalar field's backreaction on the spacetime geometry is negligible, the conditions $(\partial_r \phi)^2 \ll M_p^2 T$ and $\epsilon \phi^2 T \ll M_p^2 T$ must be satisfied.
Expressed via dimensionless model parameters (\ref{eqn9}), these constraints become
\begin{eqnarray}\label{eqn14}
\frac{\phi^2_{\text{in}}(0)}{M_p^2} \ll \frac{R_\odot^2 T}{(d\hat{\phi}/dr^\prime)^2},
\end{eqnarray}
and
\begin{eqnarray}\label{eqn15}
\frac{\phi^2_{\text{in}}(0)}{M_p^2} \ll \frac{1}{\epsilon \hat{\phi}^2(r^\prime)}.
\end{eqnarray}
These final forms provide direct upper bounds on the central field value, $\phi_{\text{in}}(0)$, relative to the reduced Planck mass, $M_p$.
The bounds depend on the solar radius squared, $R_\odot^2$, and the local geometric term, namely the torsion scalar $T$.

\begin{figure}[H]
	\begin{subfigure}{.5\textwidth}
		\centering
		\includegraphics[scale=0.5]{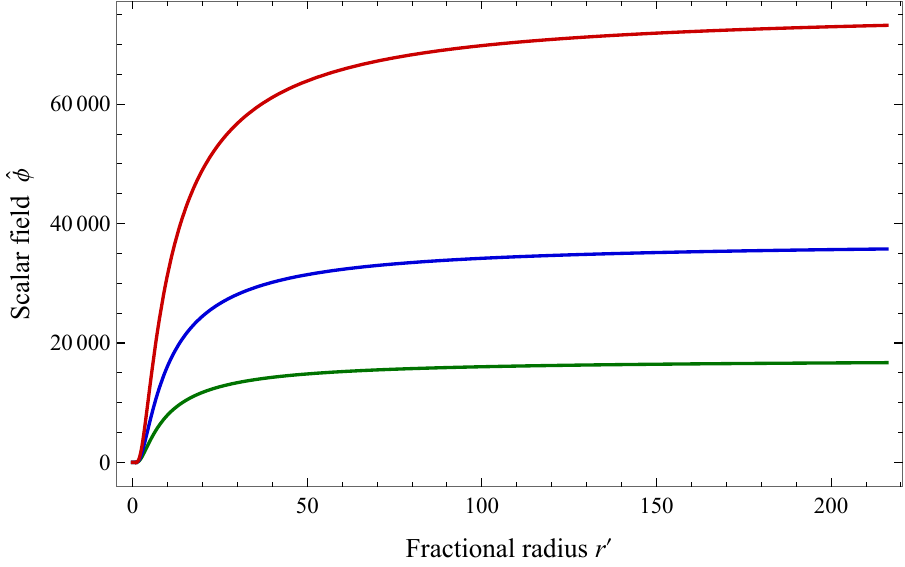}
		\label{fig1-a}
		\caption{\footnotesize{}}
	\end{subfigure}
	\begin{subfigure}{.5\textwidth}
		\centering
		\includegraphics[scale=0.5]{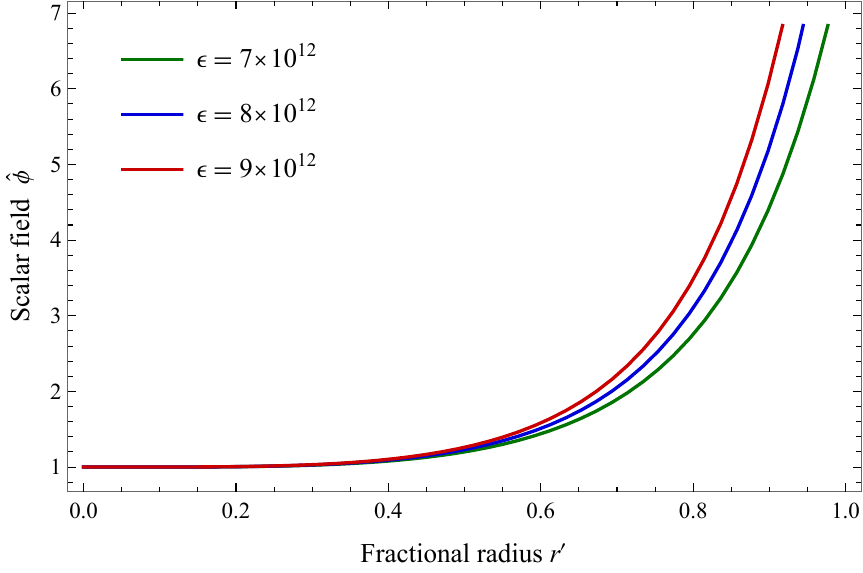}
		\label{fig1-b}
		\caption{\footnotesize{}}
	\end{subfigure}
	\caption{\footnotesize{The dimensionless scalar field profile as a function of the dimensionless fractional radius $r^\prime$.
			These cases correspond to $ r_s^{\prime}= 4.17\times 10^{-6}$ (the solar case) and $\epsilon \in \{7 \times 10^{12}, 8 \times 10^{12}, 9\times 10^{12}\}$.
			At large distances, the scalar field exhibits smooth behavior, see panel (a).
			Panel (b) is drawn to illuminate the dependence of $\hat{\phi}(r^\prime)$ on the parameters inside the object.}}
	\label{fig1}
\end{figure}

Figure \ref{fig1} illustrates the behavior of a scalar field $\hat{\phi}$ as a function of the fractional radius $R$, likely in the context of the teleparallel model of gravity.
The plot shows how the field evolves from the center of a spherical object (such as a star or planet) outward toward its surface and beyond for different values of the parameter $\epsilon$.
Notably, the numerical results indicate that the field exhibits an overlapping for the different values of $\epsilon$.
This suggests that, within the considered parameter range, the specific value of the coupling $\epsilon$ has a negligible influence on the resulting scalar field configuration (inside the sphere).
As depicted in Fig. \ref{fig1} and described by Eqs. (\ref{eqn14}) and (\ref{eqn15}), a quantitative bound may be imposed on the central scalar field value $\phi_{\text{in}}(0)$.
This derivation utilizes the field solutions obtained from Eqs. (\ref{eqn10}) and (\ref{eqn13}), which produce the respective ranges $\hat{\phi}_{\text{in}} \sim 1-10$ and $d\hat{\phi}_{\text{in}}/dr^\prime \sim 0-10^2$ from the solar interior to its surface.
Considering the characteristic parameters $\epsilon \sim 10^{12}-10^{13}$, $|T_{\text{in}}| \sim 10^{-48}-10^{-42}~\text{eV}^2$, and the solar radius $R_\odot \simeq 8.6 \times 10^{42} M_p^{-1}$, both Eqs. (\ref{eqn14}) and (\ref{eqn15}) impose the condition $\phi_{\text{in}}(0) \ll 10^{-7} M_p$.
Consequently, any backreaction on the metric remains negligible.

\begin{figure}[H]
	\centering
	\includegraphics[scale=0.5]{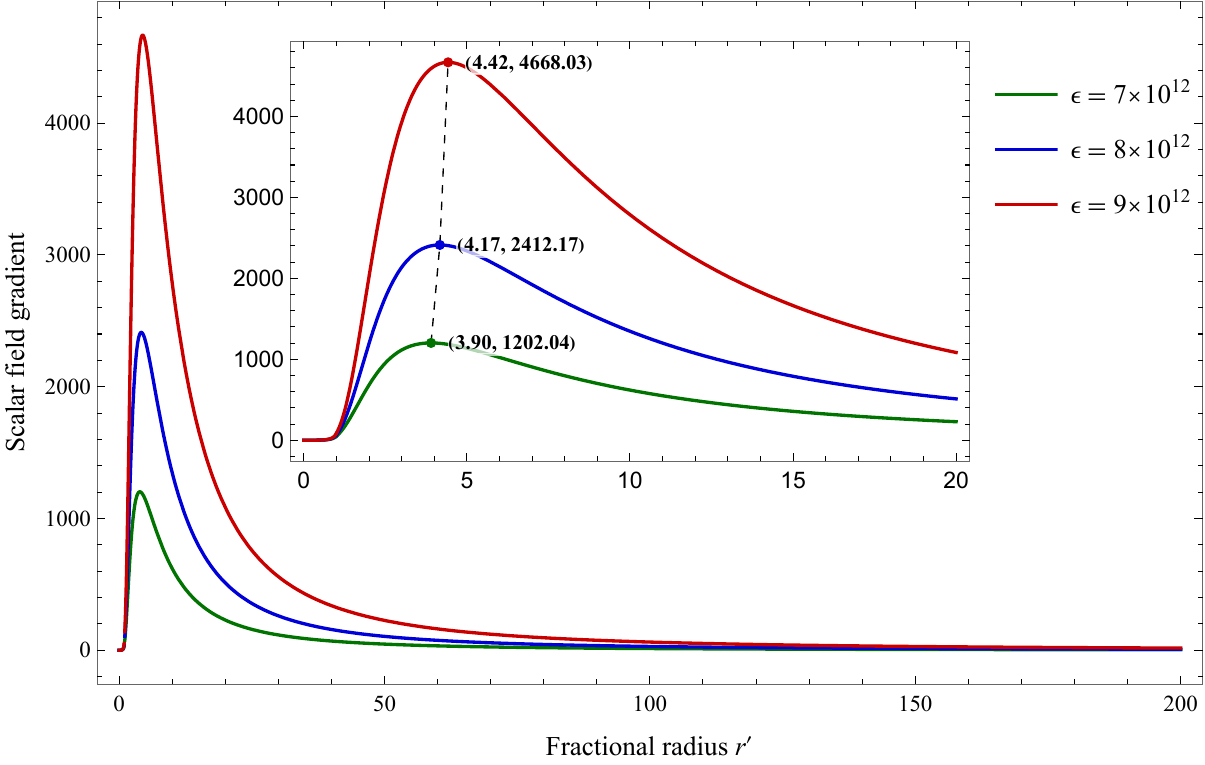}
	\caption{\footnotesize{This figure shows the scalar field gradient as a function of fractional radius $r^\prime$, showing that peaks outside the sphere increase with larger values of the scalar field-torsion coupling parameter $\epsilon$ and occur at greater distances from the sphere's surface.}}
	\label{fig2}
\end{figure}

The observations shown in Fig. \ref{fig1} indicate that each of the plotted profiles converges to finite asymptotic values outside the sphere, with the magnitude of these values increasing monotonically as the parameter $\epsilon$ grows.
Furthermore, an increase in the coupling parameter $\epsilon$ is associated with a greater difference in the scalar field values, $\hat{\phi}(r^\prime)$, manifesting as peak positions situated outside the surface of the sphere.
This trend is further illustrated in Fig. \ref{fig2}, which displays the scalar field gradient $\frac{d\hat{\phi}(r^\prime)}{dr^\prime}$ as a function of the fractional radius $r^\prime$, both in the interior and exterior of the sphere, across a range of $\epsilon$ values.
These findings imply that the influence of the scalar field-torsion coupling permeates more deeply into the outer region, potentially shaping the scalar field dynamics at progressively larger distances from the sphere.

\section{Torsion Effects on Neutrino Oscillations}\label{sec3}
A fundamental prerequisite for neutrino oscillation is the existence of non-zero neutrino mass; in the massless limit, flavor mixing is precluded.
While interactions with a medium, as first elucidated by Wolfenstein \cite{Wolf,Wolf-MSW}, can induce effective mass differences between neutrino flavors, a bare mass term remains a necessary condition.
In the context of curved spacetime, it has been proposed that geometry itself may mediate an additional interaction with ambient electrons, thereby contributing to the neutrino Hamiltonian \cite{MohseniSadjadi-Yazdani1,MohseniSadjadi-Yazdani2,MohseniSadjadi-Yazdani3,MohseniSadjadi-Yazdani4,MohseniSadjadi-Yazdani5}.

Generally, the present work explores this phenomenon within the framework of teleparallel gravity \cite{Capolupo1,Capolupo2,Panda,Teleparallel-2,Teleparallel-1,Teleparallel-3,Teleparallel-4,Teleparallel-5,Teleparallel-6,Teleparallel-7,Weitzenbock}.
In this formulation, gravitation is characterized by spacetime torsion, with the Weitzenb$\ddot{\text{o}}$ck connection—which is curvature-less but possesses torsion—supplanting the torsion-free Levi-Civita connection of General Relativity \cite{Weitzenbock}.

So, this study examines the distinct effect of torsion on neutrino oscillations within a curvature-free spacetime.
A conventional approach to directly probing torsion involves computing its coupling to the axial-vector field $A^\mu$, which includes torsion contributions \cite{Cardall,Buoninfante}.
Nevertheless, under spherically symmetric conditions—frequently encountered in astrophysical contexts—all components of this axial vector vanish identically, yielding $A^\mu = 0$ \cite{Piriz}.
The covariant derivative $D_\mu = \partial_\mu + \Gamma_\mu$, where $\Gamma_\mu = \tfrac{1}{8} [\gamma^{\hat{a}},\gamma^{\hat{b}}] e^\nu_{\hat{a}} e_{\hat{b} \nu ; \mu}$, is such that its contraction $\gamma^{\hat{a}} e^\mu_{\hat{a}} \Gamma_\mu$ is directly proportional to $A_\mu$ \cite{Cardall,Buoninfante}.
Hence, in the Dirac equation, $D_\mu$ reduces to the ordinary partial derivative $\partial_\mu$, and the gamma matrices assume their standard special-relativistic form.
Accordingly, the present analysis adopts an indirect approach: torsion is considered to affect neutrino oscillations through its contribution to the background scalar field $\hat{\phi}$, specifically via the torsion scalar $T$ and effective neutrino mass term.

For a neutrino propagating through a background of fermionic matter at standard densities within this spacetime, the effective Hamiltonian governing three-flavor oscillations can be expressed as
\begin{eqnarray}\label{eqn16}
\mathcal{H}_{\text{eff.}} = \mathcal{H}_{\text{vac}} + \mathcal{H}_{\text{mat}}.
\end{eqnarray}
The temporal evolution of the neutrino mass eigenstates, incorporating the effects of the torsion-coupled scalar field, is described by the Schr$\ddot{\text{o}}$dinger-like equation
\begin{eqnarray}\label{eqn17}
	i\frac{d}{R_\odot dr^\prime}
	\begin{pmatrix}
		{\nu_1}\\
		{\nu_2} \\
		\nu_3
	\end{pmatrix}
	=
	\left[\frac{1}{2E_\nu}
	\begin{pmatrix}
		m_1^2(\phi) & 0 & 0 \\
		0 & m_2^2(\phi) & 0 \\
		0 & 0 & m_3^2(\phi)
	\end{pmatrix}
	+ U
	\begin{pmatrix}
		\sqrt{2} G_F n_e & 0 & 0 \\
		0 & 0 & 0 \\
		0 & 0 & 0
	\end{pmatrix}
	U^{\dagger}\right]
	\begin{pmatrix}
		\nu_1 \\
		\nu_2 \\
		\nu_3
	\end{pmatrix}
	.
\end{eqnarray}
Here, $n_e$ denotes the electron number density, $G_F$ is the weak Fermi constant, and $U$ represents the Pontecorvo-Maki-Nakagawa-Sakata (PMNS) leptonic mixing matrix.
The PMNS matrix is parameterized by three mixing angles—$\theta_{12}$, $\theta_{13}$, and $\theta_{23}$—and one CP-violating phase, $\delta_{CP}$.
Crucially, the masses of the neutrino eigenstates, $m_i(\phi)$'s, are functions of the background scalar field $\phi$, as dictated by the scalar-neutrino coupling term in the action (see Eq.(\ref{eqn1})).
To isolate the contribution from the torsional coupling, we perform a series of manipulations on the evolution equation.
This allows us to derive effective, torsion-induced modifications to the neutrino mass-squared differences and mixing parameters within a material medium.

\subsection{Modified Oscillations in Vacuum}\label{subsec3-1}
For a wide array of experimental setups, the intricate three-flavor oscillation pattern can be accurately modeled using a simpler two-neutrino approximation \cite{Agarwalla,Martinez-Mirave}.
In this context, the vacuum Hamiltonian for two flavors is modified. After rotating to the flavor basis via the PMNS matrix, the Hamiltonian reads
\begin{eqnarray}\label{eqn18}
	\mathcal{H}_{\text{vac}} = \frac{\Delta m_{21}^2(\phi)}{4E_\nu}
	\begin{pmatrix}
		-\cos 2\theta & \sin 2\theta \\
		\sin 2\theta & \cos 2\theta
	\end{pmatrix}.
\end{eqnarray}
In such a scheme, the transition probability between two flavors simplifies considerably as

\begin{eqnarray}\label{eqn19}
	P_{\alpha \beta} = \sin^2 (2\theta) \sin^2 \left( \frac{\Phi_{21}}{2} \right).
\end{eqnarray}
The core of this probability is the phase difference $\Phi_{21}$, which accumulates along the neutrino's path. Our derivation begins with an expression for this phase, grounded in the Dirac equation in the presence of torsion, as in our previous works \cite{MohseniSadjadi-Yazdani1, MohseniSadjadi-Yazdani2, MohseniSadjadi-Yazdani3,MohseniSadjadi-Yazdani4,MohseniSadjadi-Yazdani5,MohseniSadjadi-Yazdani6}.
This takes the form
\begin{eqnarray}\label{eqn20}
\begin{split}
\Phi_{ij} &= \int_{x_A}^{x_B} p_\mu dx^\mu \\& = \int_{r_A}^{r_B} \frac{\Delta m_{ij}^2(\phi)}{2E_\nu} dr = \frac{R_\odot \Delta \kappa_{21}^{\prime 2}}{2E_\nu} \int_{r_A^\prime}^{r_B^\prime} \hat{\phi}^4(r^\prime) dr^\prime,
\end{split}
\end{eqnarray}
where $r_A^\prime$ and $r_B^\prime$ denote the radial positions of the neutrino source and detector, respectively.
Drawing from the neutrino mass generation mechanism in Eq.(\ref{eqn1}), we posit $\Delta m_{21}^2(\hat{\phi}) = \Delta\kappa_{21}^{\prime 2} \hat{\phi}^4(r)$.
Here, a rescaled coupling $\kappa_{i}^{\prime} \equiv \kappa_i \phi^2_{\text{in}}(0)$ (in eV) is defined, which absorbs the unknown central value of the scalar field, $\phi_{\text{in}}(0)$.
This rescaling highlights a key challenge: without an independent constraint on $\phi_{\text{in}}(0)$, the absolute scale of neutrino masses remains elusive.
Consequently, our subsequent comparison with solar neutrino data will aim to directly constrain the combined parameter $\Delta \kappa_{21}^{\prime 2}$.

Under the weak-field approximation, the condition $r_s^\prime \ll 1$ may be justifiably applied to the solar case.
Adopting this approximation and using Eqs. (\ref{eqn10}) and (\ref{eqn13}), for the region where both $r_A^\prime < 1$ and $r_B^\prime < 1$, the expression for $\Phi_{ij}$ reduces to
\begin{eqnarray}\label{eqn21}
\Phi_{ij} = \frac{R_\odot \Delta \kappa_{21}^{\prime 2}}{2E_\nu} \left[r_B^\prime - r_A^\prime + \frac{r_s^{\prime 2} \epsilon}{50} \left(r_B^{\prime 5} - r_A^{\prime 5}\right)\right].
\end{eqnarray}
In the exterior regime, where $r_A^\prime > 1$ and $r_B^\prime > 1$, the corresponding expression becomes
\begin{eqnarray}\label{eqn22}
\Phi_{ij} = \frac{R_\odot \Delta \kappa_{21}^{\prime 2}}{2E_\nu} \left[-\frac{12}{5} r_s^{\prime 2} \epsilon  \ln\left(\frac{r_B^\prime}{r_A^\prime}\right) + \frac{(r_B^\prime - r_A^\prime) \left(r_s^{\prime 2} \epsilon  (3 r_A^\prime r_B^\prime+2) + 2 r_A^\prime r_B^\prime\right)}{2 r_A^\prime r_B^\prime}\right].
\end{eqnarray}
For the mixed case, where $r_A^\prime < 1$ and $r_B^\prime > 1$, the expression for $\Phi_{ij}$ takes the following form:
\begin{eqnarray}\label{eqn23}
\Phi_{ij} = \frac{R_\odot \Delta \kappa_{21}^{\prime 2}}{2E_\nu} \left[r_B^\prime - r_A^\prime +  \frac{r_s^{\prime 2} \epsilon  \left(-\left(r_A^{\prime 5} + 24\right) r_B^\prime + 75 r_B^{\prime 2} - 50\right)}{50 r_B^\prime} -\frac{12}{5} r_s^{\prime 2} \epsilon  \log (r_B^\prime)\right].
\end{eqnarray}
In each of the preceding cases, the phase difference converges to the conventional expression in the limit $\epsilon \to 0$.
Within the framework of the teleparallel model of gravity considered in this study, the dimensionless product $r_s^{\prime 2} \epsilon$ plays a critical role in quantifying the departure from standard neutrino oscillation physics.
For the Sun, $r_s^\prime \sim 10^{-5}$ is small, consistent with the weak-field approximation.
However, as constrained by solar neutrino data (see section \ref{sec4}) and screening mechanisms \cite{sym}, the coupling parameter $\epsilon$ is required to be large, typically $\epsilon \sim 10^{12} - 10^{13}$.
Consequently, their product $r_s^{\prime 2} \epsilon$ becomes of order $\mathcal{O}(10^2)$.
This non-standard combination appears naturally in the expressions of modified neutrino phase differences derived above.
Thus, despite the smallness of the gravitational field, the enhanced coupling may amplify the torsional effect, making it potentially detectable via future observations.

\subsection{Torsion-Induced MSW Effect}\label{subsec3-2}
In the conventional MSW mechanism \cite{Wolf-MSW, Mikh-Smir-MSW}, resonant flavor conversion originates from the interplay between the vacuum mass splitting and the matter-induced potential. Within the framework of torsion-coupled gravity, an additional degree of freedom emerges in the form of the scalar field $\hat{\phi}(r^\prime)$.
This field modifies exclusively the vacuum contribution—scaling it by a factor of $\hat{\phi}^4(r^\prime)$—while leaving the matter potential unchanged.
Consequently, all flavor conversion dynamics are governed by a single dimensionless parameter
\begin{eqnarray}\label{eqn24}
A = \pm \frac{2\sqrt{2} G_F n_e E_\nu}{\Delta \kappa_{21}^{\prime 2} \hat{\phi}^4(r_B^\prime)},
\end{eqnarray}
where the upper (lower) sign applies to neutrinos (antineutrinos).
In terms of this parameter, the effective Hamiltonian (\ref{eqn16}) takes the form
\begin{eqnarray}\label{eqn25}
\mathcal{H}_{\text{eff.}} = \frac{\Delta \kappa_{21}^{\prime 2} \hat{\phi}^4(r_B^\prime)}{4E_\nu}
\begin{pmatrix}
	-\cos 2\theta + A & \sin 2\theta \\
	\sin 2\theta & \cos 2\theta - A
\end{pmatrix},
\end{eqnarray}
which reverts to the standard MSW Hamiltonian in the limit of a constant scalar field ($\hat{\phi} \to \text{constant}$).
\paragraph{Three distinct regimes inside the Sun.}
Owing to the radial dependence of $\hat{\phi}(r^\prime)$ and the electron density $n_e(r^\prime)$, a neutrino propagating through the Sun experiences three well-separated regimes:
\begin{itemize}
\item \textbf{Solar core ($\hat{\phi}_{\text{in}} \to 1$, high $n_e$):}
In this region $|A| \gg 1$, so the diagonal entries of $\mathcal{H}_{\text{eff}}$ dominate, leading to strong suppression of mixing.
The torsion-induced effects remain effectively masked by the high matter density.
\item \textbf{Resonance layer ($A = \cos 2\theta$):} Here the diagonal elements cancel.
Diagonalizing the Hamiltonian yields the matter-modified mixing angle
\begin{eqnarray}\label{eqn26}
\tan 2\theta_m = \frac{\sin 2\theta}{\cos 2\theta - A},
\end{eqnarray}
which attains $\pi/4$ exactly at resonance.
Also, the corresponding effective mass-squared difference is
\begin{eqnarray}\label{eqn27}
\Delta m^2_m = \left[ \Delta \kappa_{21}^{\prime 2} \hat{\phi}^4(r_B^\prime;\epsilon) \right] \sqrt{1 + A^2 - 2A \cos 2\theta}.
\end{eqnarray}
\item \textbf{Outer solar layers ($\hat{\phi} \to \hat{\phi}_\infty$, low $n_e$):}
In this regime $A \to 0$, so the Hamiltonian reduces to its torsion-modified vacuum form, and oscillations proceed as in vacuum (see subsection \ref{subsec3-1}).
\end{itemize}

\paragraph{Adiabatic regime and survival probability.}
For high-energy neutrinos, such as solar $^8$B neutrinos whose energies exceed the resonance value, the transition between different propagation regimes occurs adiabatically. Following the method outlined in Ref.~\cite{MohseniSadjadi-Yazdani3}, the survival probability of electron neutrinos is expressed as
\begin{eqnarray}\label{eqn28}
P_{ee}(E_\nu) = \frac{1}{2} \left[ 1 + \cos 2\theta_m \cos 2\theta \right].
\end{eqnarray}
Conversely, for lower-energy neutrinos (e.g., $pp$ and $^7$Be neutrinos), $P_{ee}$ varying smoothly between the vacuum-dominated limit and the matter-dominated behavior described previously.

Equation (\ref{eqn28}) for the electron neutrino survival probability assumes adiabatic neutrino propagation. To validate this assumption, the adiabaticity parameter $\gamma$ must satisfy $\gamma \gg 1$ along the entire neutrino trajectory. This parameter is defined as \cite{Strumia}
\begin{eqnarray}\label{eqn29}
\gamma(r^\prime) \equiv \frac{R_\odot \Delta m^2_M(r^\prime)/4E_\nu}{|d\theta_M/dr^\prime|}.
\end{eqnarray}
Employing the modified mixing angle from Eq. (\ref{eqn26}), the derivative in the denominator takes the form $\frac{d\theta_M}{dr^\prime} = \frac{\sin 2\theta }{2 [ 1 + A^2 - 2A \cos 2\theta ] } \frac{dA}{dr^\prime}$, yielding the general adiabaticity expression
\begin{eqnarray}\label{eqn30}
\gamma(r^\prime) = \frac{ R_\odot \Delta \kappa_{21}^{\prime 2} \hat{\phi}^4(r_B^\prime)\left[1 + A^2 - 2 A\cos2\theta\right]^{\frac{3}{2}}}{2E_\nu \sin 2\theta~|\frac{d A}{dr^\prime}|}.
\end{eqnarray}
The most stringent adiabaticity condition occurs at the resonance point, where $A_{\text{res.}} \to \cos 2\theta$. Under this resonance condition, the expression simplifies to
\begin{eqnarray}\label{eqn31}
\gamma_{\text{res.}} = \tilde{\gamma}_\text{{res.}} \left[\frac{\sin^2 2\theta}{2 \pi \cos2\theta}\right],
\end{eqnarray}
with $\tilde{\gamma}_{\text{res.}} = \frac{\pi R_\odot \Delta \kappa_{21}^{\prime 2} \hat{\phi}^4(r_B^\prime)/E_\nu}{|d\ln A/dr^\prime|_{\text{res.}}}$.
The radial derivative $d \ln A/dr^\prime$ depends on the electron number density profile $n_e(r^\prime)$. Adopting the established approximation $n_e(r^\prime) = 245 N_A/\text{cm}^3 \times \exp \left(-10.54~r^\prime \right)$ \cite{Strumia,Bahcall}, where $N_A = 6.022 \times 10^{23} \text{mol}^{-1}$ is Avogadro's constant, the parameter $\tilde{\gamma}_{\text{res.}}$ can be approximated by
\begin{eqnarray}\label{eqn32}
\tilde{\gamma}_{\text{res.}} \approx \frac{\Delta \kappa_{21}^{\prime 2} \hat{\phi}^4(r_B^\prime) / E_\nu}{10^{-9} \text{eV}^2/\text{MeV}}.
\end{eqnarray}
Within the framework of the large mixing angle (LMA) solution to the solar neutrino problem, where $\Delta m^2_{21} (\text{on Earth}) = \Delta \kappa_{21}^{\prime 2} \hat{\phi}^4(r_B^\prime) \sim 7\times 10^{-5} \text{eV}^2$, the resonance adiabaticity parameter satisfies $\gamma_{\text{res.}} \gtrsim 10^3$ for $^8$B neutrinos with energies near $10~\text{MeV}$. Consequently, the adiabatic expression in Eq. (\ref{eqn28}) is fully justified for the model parameters and astrophysical environments considered in this study.

\section{Bounds on Model Parameters}\label{sec4}
Model constraints are obtained via a least-squares method, by comparing the theoretical electron neutrino survival probability $P_{ee}^{\text{th}}(E_\nu)$ with experimentally derived values from solar neutrino data.
The dataset incorporates measurements from Kamiokande \cite{Kamiokande}, SK phases I–IV \cite{Super-Kamiokande,SK-I,SK-II,SK-III}, SNO phases I–III (including CC and $\nu e$ elastic scattering channels) \cite{SNOEx,SNO-II,SNO-III}, and Borexino \cite{Borexino-Data,BOREXINO}, as summarized in the second column of Tables \ref{table1} and \ref{table2}.
For each experiment, the observed survival probability $P_i^{\text{obs}}$ is derived from the reported neutrino flux $\Phi_i$, normalized to the Standard Solar Model (SSM) [BPS08(GS)] prediction for the $^8$B flux, $\Phi_{\text{SSM}} = (5.94 \pm 0.65) \times 10^6$ cm$^{-2}$ s$^{-1}$ \cite{Pena-Garay,PDG}, via $P_i^{\text{obs}} = \Phi_i / \Phi_{\text{SSM}}$.
The corresponding uncertainty $\sigma_i$ incorporates both experimental and SSM-related errors propagated accordingly.

To compute $P_i^{\text{th}}$ for a given parameter set $\{\epsilon,\Delta\kappa_{21}^{\prime 2}\}$, we identify $r_A^\prime$ and $r_B^\prime$ as the dimensionless neutrino production and detection radii, normalized to the solar radius $R_\odot$.
For each solar neutrino component, the production point $r_A^\prime$ is taken to be the corresponding flux-weighted mean production radius predicted by SSM \cite{Bahcall}.
In particular, $^8$B neutrinos are assumed to originate at $r_A^\prime \simeq 0.05$, while the appropriate SSM production radii can be used for the $pp$, $pep$, and ${}^{7}$Be neutrino components ($r_A^\prime \sim 0.1 - 0.3$) measured by Borexino \cite{BOREXINO}.
The detection point is taken to be the Earth, corresponding to $r_B^\prime = 1\text{AU}/R_\odot\simeq215$, where the scalar field reaches its asymptotic value $\hat{\phi}_\infty$.
Consequently, the effective terrestrial mass-squared difference is $\Delta m_{21}^2({\text{Earth}}) = \Delta\kappa_{21}^{\prime 2}\hat{\phi}_\infty^4$.
For each parameter pair $\{\epsilon,\Delta\kappa_{21}^{\prime 2}\}$, the scalar field profile $\hat{\phi}(r^\prime)$ is first computed from Eqs. (\ref{eqn10}) and (\ref{eqn13}).
This profile is then used to evaluate the theoretical survival probability $P_{ee}^{\text{th}}(E_\nu)$ through Eqs.~(\ref{eqn26})--(\ref{eqn28}).
$P_{ee}^{\text{th}}(E_\nu)$ is evaluated at the characteristic energy of the corresponding neutrino component.
For the $^8$B neutrino analyses (Kamiokande, Super-Kamiokande, and SNO), we adopt $E_\nu \simeq 10\,{\text{MeV}}$, while for Borexino, the survival probability can be evaluated at the central energies of the corresponding measured solar neutrino components ($pp$, $pep$, and $^7$Be).

\begin{figure}[H]
	\centering
	\includegraphics[scale=0.40]{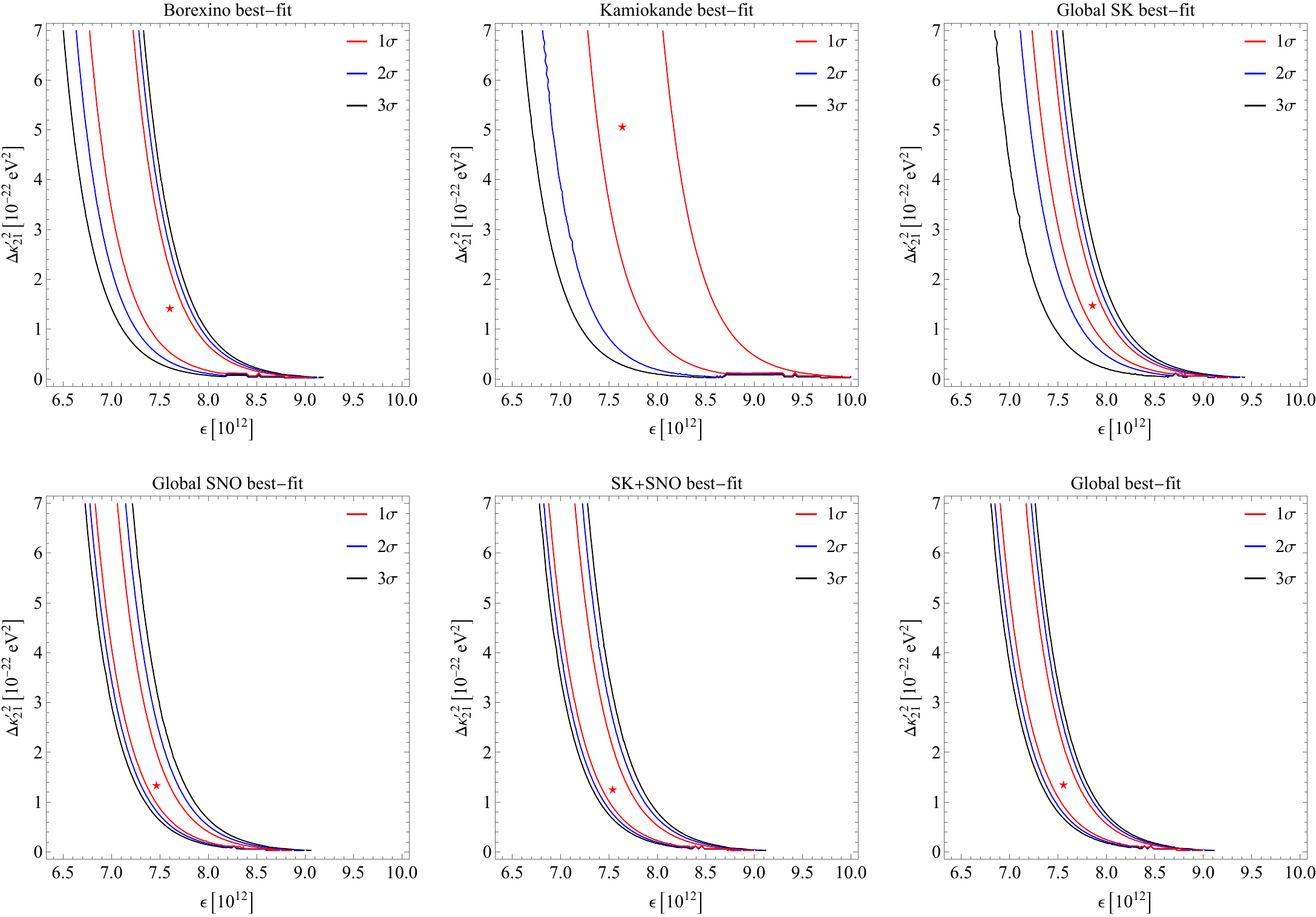}
	\caption{\footnotesize{The analysis of confidence regions (1$\sigma$, 2$\sigma$, and 3$\sigma$) for the model parameters $\epsilon$ (dimensionless) and $\Delta \kappa^{\prime 2}_{21}$.
			The best-fit values are shown by asterisks.}}
	\label{fig3}
\end{figure}

The agreement between theoretical predictions and experimental data is evaluated using a $\chi^2$ statistic, expressed as
\begin{eqnarray}\label{eqn33}
\chi^2(\epsilon, \Delta\kappa^{\prime 2}_{21}) = \sum_{i} \frac{\left(P_i^{\text{obs}} - P_i^{\text{th}}(\epsilon, \Delta\kappa^{\prime 2}_{21})\right)^2}{\sigma_i^2}.
\end{eqnarray}
For each experiment, an individual $\chi^2$ function is constructed, while combined analyses are obtained by summing the corresponding contributions.
The parameters $\epsilon$ and $\Delta\kappa_{21}^{\prime 2}$ are treated as independent free parameters and are varied simultaneously to determine the global minimum of the two-dimensional $\chi^2$ function.
No external priors or fixed values are imposed on either parameter when constraining the other.
The individual confidence intervals reported in Tables~\ref{table1} and \ref{table2} are obtained by profiling over the remaining parameter, while the joint constraints are represented by the two-dimensional $\Delta\chi^2$ confidence contours.
All uncertainties are treated as uncorrelated, and correlated systematic effects are neglected in the present analysis.
The resulting $1\sigma$, $2\sigma$, and $3\sigma$ allowed regions are shown in Fig.~\ref{fig3}, while the corresponding best-fit values and confidence intervals are reported in Tables~\ref{table1} and \ref{table2} for both individual experiments and combined datasets.

Naturally, these findings may carry significant implications for our understanding of cosmology and gravitational physics, as they place constraints on extensions of the Standard Model—such as TEGR.
Table \ref{table1} compiles measurements of the $^8$B solar neutrino flux and the associated parameter $\epsilon$, which plays a key role in characterizing scalar field behavior (see figure \ref{fig1}).
For each dataset, the table lists the best-fit value of $\epsilon$ along with the corresponding $2\sigma$ and $3\sigma$ confidence intervals.

Overall, the best-fit estimates for $\epsilon$ remain remarkably stable across different experiments, clustering around $7.654 \times 10^{12}$. The values span from a low of $7.446 \times 10^{12}$ (from the SNO-Phase III $\nu e$ channel) to a high of $7.865 \times 10^{12}$ (SK-II). When all data are combined in a global fit, the resulting best-fit value is $\epsilon = 7.560 \times 10^{12}$, offering the most statistically robust determination of the parameter.

The reported $2\sigma$ and $3\sigma$ intervals offer a detailed view of the measurement uncertainties.
The combined multi-experiment analyses—global SK ($2\sigma$: $[7.691, 7.978] \times 10^{12}$, $3\sigma$: $[7.463, 8.040] \times 10^{12}$), global SNO ($2\sigma$: $[7.335, 7.648] \times 10^{12}$, $3\sigma$: $[7.280, 7.732] \times 10^{12}$), the SK+SNO combination ($2\sigma$: $[7.403, 7.757] \times 10^{12}$, $3\sigma$: $[7.352, 7.825] \times 10^{12}$), and the overall global fit ($2\sigma$: $[7.402, 7.739] \times 10^{12}$, $3\sigma$: $[7.350, 7.789] \times 10^{12}$)—consistently yield the tightest confidence regions.
This reduction in uncertainty highlights the statistical benefit of integrating multiple datasets.

Furthermore, as explained in the discussion around Eq.~(\ref{eqn23}), the theoretical treatment of torsion-induced modifications intrinsically points to a characteristic scale for the coupling parameter, specifically within the range $\epsilon \sim 10^{12}$--$10^{13}$.
Remarkably, all best-fit values obtained from the solar neutrino analysis lie within this expected interval.
For comparison, in a limiting case of the screening models discussed in Ref.~\cite{sym,Sadjadi-Khosravi}, where the coupling term takes the form $\epsilon R\phi^2/2$ in the presence of Ricci curvature, the corresponding parameter is required to satisfy $\epsilon > 10^{6}$.
The values obtained in the present analysis are therefore fully compatible with this lower bound while providing a substantially more restrictive determination of the coupling strength.

\begin{table}[H]
	\begin{center}
		\tiny
		\caption{\footnotesize{Results from solar neutrino experiments regarding $^8$B neutrino flux, the best-fit values and corresponding 2$\sigma$ and 3$\sigma$ confidence intervals have been shown.
				The best-fit values are obtained around $7.654 \times 10^{12}$.
		}}
		\label{table1}
		\begin{tabular}{|c|c|c|c|c|}
			\hline
			\hline
			Experiment & $\nu_e$ flux [$10^6/\text{cm}^2 \text{s}$] & $\epsilon$ best-fit $[10^{12}]$ &  2$\sigma$ range $[10^{12}]$ & 3$\sigma$ range $[10^{12}]$ \\
			\hline \hline
			Kamiokande \cite{Kamiokande} & $2.80 \pm 0.19$ &$7.643$ & $>7.251$ & $>6.782$ \\ \hline
			SK-I \cite{SK-I} & $2.38 \pm 0.02$ &$7.833$ & $[7.246 , 8.090]$ & $[7.065 , 8.301]$ \\
			SK-II \cite{SK-II} & $2.41 \pm 0.05$ &$7.865$ & $[7.283 , 8.135]$ & $[7.089 , 8.383]$ \\
			SK-III \cite{SK-III} & $2.32 \pm 0.04$ &$7.806$ & $[7.209 , 8.056]$ & $[7.050 , 8.235]$\\
			SK-IV \cite{Super-Kamiokande} & $2.31 \pm 0.02$ &$7.820$ & $[7.232 , 8.060]$ & $[7.069 , 8.224]$ \\
			Global SK (SK I-IV) & &$7.857$ & $[7.691 , 7.978]$ & $[7.463 , 8.040]$ \\ \hline
			SNO-Phase I (CC) \cite{SNOEx} & $1.76^{+0.06}_{-0.05}$ & $7.454$ & $[7.179 , 7.796]$ & $[7.051 , 7.903]$ \\
			SNO-Phase I ($\nu e$) \cite{SNOEx} & $2.39^{+0.24}_{-0.23}$ & $7.858$ & $[7.195 , 8.244]$ & $[7.053 , 9.208]$ \\
			SNO-Phase II (CC) \cite{SNO-II} & $1.68 \pm 0.06$ & $7.452$ & $[7.187 , 7.773]$ & $[7.059 , 7.881]$ \\
			SNO-Phase II ($\nu e$) \cite{SNO-II} & $2.35 \pm 0.22$ & $7.798$ & $[7.149 , 8.155]$ & $[7.009 , 8.661]$ \\
			SNO-Phase III (CC) \cite{SNO-III} & $1.67^{+0.05}_{-0.04}$ & $7.452$ & $[7.189 , 7.766]$ & $[7.062 , 7.873]$ \\
			SNO-Phase III ($\nu e$) \cite{SNO-III} & $1.77^{+0.24}_{-0.21}$ & $7.446$ & $[7.161 , 7.867]$ & $[7.032 , 8.014]$ \\
			Global SNO &  & $7.469$ & $[7.335 , 7.648]$ & $[7.280 , 7.732]$ \\ \hline
			SK+SNO &  & $7.543$ & $[7.403 , 7.757$ & $[7.352 , 7.825]$ \\ \hline
			Borexino  \cite{Borexino-Data,BOREXINO} & $2.57^{+0.17}_{-0.18}$ & $7.603$ & $[7.209 , 7.783]$ & $[7.071 , 7.839]$ \\ \hline
			Global fit & & $7.560$ & $[7.402 , 7.739]$ & $[7.350 , 7.789]$ \\
			\hline
		\end{tabular}
	\end{center}
\end{table}
To quantify the statistical preference for a non-zero coupling $\epsilon$, we compare the best-fit $\chi^2$ of our model (with both $\epsilon$ and $\Delta\kappa_{21}^2$ free) against the standard LMA-MSW limit ($\epsilon \to 0$).
For the global dataset, we find $\Delta\chi^2 = \chi^2_{\epsilon=0} - \chi^2_{\mathrm{free}} = 8.89$ for one additional parameter, corresponding to a preference of approximately $2.98 \sigma$ for the torsion-induced modification.
The improvement in fit quality originates from the additional freedom provided by the torsion-induced modification of the scalar field profile, which allows a slightly better description of the observed solar neutrino survival probabilities than the standard LMA-MSW limit.
This indicates a moderate hint of new physics, though not yet at the $3\sigma$ level typically required to claim evidence.
Since no external priors are imposed on $\epsilon$, the preference for a non-zero coupling arises entirely from the solar neutrino data and is not driven by assumptions built into the fitting procedure.
The robustness of this result was verified through leave-one-out tests, which confirm that the preference persists when any single experiment is removed, albeit with reduced significance.

Table \ref{table2} summarizes the measured values of the neutrino coupling difference $\Delta \kappa^{\prime 2}_{21}$, given in units of $10^{-22} \text{eV}^2$. These estimates are derived from fits to both individual neutrino oscillation datasets and a combined global analysis that incorporates all available flux measurements.
The values vary between a minimum of roughly $1.250 \times 10^{-22} \text{eV}^2$ (for the SK+SNO combination) and a maximum of $5.053 \times 10^{-22} \text{eV}^2$ (from Kamiokande).
The global fit returns a central value of $1.346 \times 10^{-22} \text{eV}^2$, pointing to broad agreement among the various experimental results.

The combined fits—such as global SK ($1.474 \times 10^{-22} \text{eV}^2$), global SNO ($1.326 \times 10^{-22} \text{eV}^2$), SK+SNO ($1.250 \times 10^{-22} \text{eV}^2$), and the overall global fit ($1.346 \times 10^{-22} \text{eV}^2$)—tend to yield tighter $2\sigma$ and $3\sigma$ confidence intervals than those from individual experiments.
This demonstrates the benefit of data aggregation in reducing statistical uncertainties.

Because $\Delta \kappa^{\prime 2}_{21}$ is directly related to the mass-squared difference $\Delta m^{2}_{21}$ between the two lightest neutrino mass eigenstates, obtaining a precise value is key to deepening our knowledge of the neutrino mass structure and the underlying mechanics of neutrino oscillations.
\begin{table}[H]
	\begin{center}
		\tiny
		\caption{\footnotesize{Determinations of $\Delta\kappa^{\prime 2}_{21}$ including best-fit values, 2$\sigma$ and 3$\sigma$ confidence intervals. }}
		\label{table2}
		\begin{tabular}{|c|c|c|c|c|}
			\hline
			\hline
			Experiment & $\nu_e$ flux [$10^6/\text{cm}^2 \text{s}$] & $\Delta \kappa^{\prime 2}_{21}$ best-fit $[10^{-22} \text{eV}^2]$ &  2$\sigma$ range $[10^{-22} \text{eV}^2]$ & 3$\sigma$ range $[10^{-22} \text{eV}^2]$ \\
			\hline \hline
			Kamiokande \cite{Kamiokande} & $2.80 \pm 0.19$ &$5.053$ & $>1.525$ & $>0.349$ \\ \hline
			SK-I \cite{SK-I} & $2.38 \pm 0.02$ &$1.639$ & $[0.276 , 3.507]$ & $[0.158 , 6.475]$ \\
			SK-II \cite{SK-II} & $2.41 \pm 0.05$ &$1.566$ & $[0.269 , 3.471]$ & $[0.147 , 7.137]$ \\
			SK-III \cite{SK-III} & $2.32 \pm 0.04$ &$1.621$ & $[0.264 , 3.399]$ & $[0.161 , 5.729]$\\
			SK-IV \cite{Super-Kamiokande} & $2.31 \pm 0.02$ &$1.548$ & $[0.259 , 3.150]$ & $[0.156 , 5.083]$ \\
			Global SK (SK I-IV) & &$1.474$ & $[0.898 , 2.111]$ & $[0.450 , 2.536]$ \\ \hline
			SNO-Phase I (CC) \cite{SNOEx} & $1.76^{+0.06}_{-0.05}$ & $1.326$ & $[0.568 , 3.724]$ & $[0.381 , 5.118]$ \\
			SNO-Phase I ($\nu e$) \cite{SNOEx} & $2.39^{+0.24}_{-0.23}$ & $1.478$ & $[0.197, 4.606]$ & $>0.127$ \\
			SNO-Phase II (CC) \cite{SNO-II} & $1.68 \pm 0.06$ & $1.278$ & $[0.565 , 3.375]$ & $[0.379 , 4.647]$ \\
			SNO-Phase II ($\nu e$) \cite{SNO-II} & $2.35 \pm 0.22$ & $1.682$ & $[0.233 , 4.835]$ & $>0.151$ \\
			SNO-Phase III (CC) \cite{SNO-III} & $1.67^{+0.05}_{-0.04}$ & $1.265$ & $[0.564 , 3.272]$ & $[0.379 , 4.490]$ \\
			SNO-Phase III ($\nu e$) \cite{SNO-III} & $1.77^{+0.24}_{-0.21}$ & $1.417$ & $[0.587 , 5.031]$ & $[0.393 , 7.778]$ \\
			Global SNO & & $1.326$ & $[0.880 , 2.277]$ & $[0.741 , 2.938]$ \\ \hline
			SK+SNO & & $1.250$ & $[0.817 , 2.383]$ & $[0.698 , 2.919]$ \\ \hline
			Borexino \cite{Borexino-Data,BOREXINO} & $2.57^{+0.17}_{-0.18}$ & $1.408$ & $[0.421 , 2.413]$ & $[0.274 , 2.856]$ \\ \hline
			Global fit & & $1.346$ & $[0.831 , 2.308]$ & $[0.708 , 2.676]$ \\
			\hline
		\end{tabular}
	\end{center}
\end{table}

The parameter $\epsilon$ does not typically appear within the standard three-neutrino oscillation framework.
Instead, it may serve as an indicator of physics BSM, potentially signaling the presence of sterile neutrinos or non-standard interaction effects \cite{Miranda}.
As such, constraints on $\epsilon$ provide valuable input for probing new physics scenarios, particularly those motivated by teleparallel gravity.
The accompanying figures illustrate how limits on this parameter correlate with another non-standard oscillation quantity, $\Delta \kappa^{\prime 2}_{21}$.
Specifically, Fig.~\ref{fig3} shows the confidence regions for the pair $\{\epsilon, \Delta \kappa^{\prime 2}_{21}\}$, offering insight into the relationship between these two extensions of the standard neutrino picture.
The correlation between $\epsilon$ and $\Delta\kappa_{21}^{\prime 2}$ observed in Fig.~\ref{fig3} arises naturally from their combined effect on the neutrino survival probability $P_{ee}(E_\nu)$.
As shown in Eqs.~(\ref{eqn26})-(\ref{eqn28}), the oscillation probability depends on the scalar field profile $\hat{\phi}(r';\epsilon)$, whose radial evolution is controlled by the torsion-scalar coupling parameter $\epsilon$, while the effective flavor conversion scale is governed by $\Delta\kappa_{21}^{\prime 2}$.
More explicitly, both parameters enter the neutrino phase difference $\Phi_{21} = (R_\odot \Delta\kappa_{21}^{\prime 2} / 2E_\nu) \int \hat{\phi}^4(r') dr'$ and terrestrial mass-squared difference $\Delta m_{21}^2 = \Delta\kappa_{21}^{\prime 2} \hat{\phi}_\infty^4$.
For fixed $\Phi_{21}$ and $\Delta m_{21}^2$, an increase in $\epsilon$ enhances the exterior scalar field $\hat{\phi}_\infty$, requiring a decrease in $\Delta\kappa_{21}^{\prime 2}$ to keep them unchanged.
This parameter degeneracy gives rise to the elongated confidence contours shown in Fig.~\ref{fig3}.
The net effect, as seen in the figure, is a negative correlation between $\epsilon$ and $\Delta\kappa_{21}^{\prime 2}$: larger values of $\epsilon$ favor smaller values of $\Delta\kappa_{21}^{\prime 2}$.
Its main implication is that the constraints on $\epsilon$ and $\Delta\kappa_{21}^{\prime 2}$ cannot be interpreted independently; instead, a simultaneous two-parameter fit is required.
The observed correlation also explains why the one-dimensional confidence intervals are broader than would be obtained if either parameter were fixed, since the uncertainty in one parameter propagates into the determination of the other.

\subsection{Sensitivity Analysis}\label{subsec4-1}
To verify the robustness of the parameter constraints, we study the effect of plausible variations in the scalar field profile and solar density distribution on the inferred values of $\epsilon$ and $\Delta\kappa_{21}^{\prime 2}$:

\subsubsection{Sensitivity to the scalar field profile}
Scalar field profile $\hat{\phi}(r^\prime)$ is obtained from Eqs.~(\ref{eqn10}) and (\ref{eqn13}) using physically motivated boundary conditions associated with screening mechanisms in scalar-tensor theories.
To assess the robustness of the results, we explicitly varied the continuity radius ($r^\prime = 1$) at the solar surface by $\pm 5\%$ and repeated the parameter extraction procedure.
This modification changes the asymptotic value $\hat{\phi}_\infty$ by less than $3\%$.
The resulting best-fit values shift by less than $1\%$ for $\epsilon$ and $< 6\%$ for $\Delta\kappa_{21}^{\prime 2}$, which is significantly smaller than the statistical uncertainties reported in Tables~\ref{table1} and \ref{table2}.
We therefore conclude that the inferred parameter constraints are stable against reasonable variations of the obtained scalar field profile.

\subsubsection{Sensitivity to the solar density model}
The solar density profile enters the analysis through the parameter $A$, Eq.(\ref{eqn24}).
For the exponential parameterization $n_e(r^\prime) = n_0 e^{-\alpha R_\odot r^\prime}$ \cite{Strumia,Bahcall} (see below relation (\ref{eqn31})), the MSW resonance radius is $r^\prime_{\text{res}} = \frac{1}{\alpha R_\odot}\ln\left(\frac{n_0}{n_{\text{res}}}\right).$
Consequently, a fractional variation of the density normalization factor produces only $\delta r^\prime_{\text{res}}=\frac1{\alpha R_\odot}\frac{\delta n_0}{n_0},$
so that even a $10\%$ modification of the density amplitude changes the resonance location by less than $1\%$ of the solar radius.
Moreover, the adiabaticity parameter depends only on the density scale height,$\gamma \propto \left|\frac{1}{R_\odot n_e}\frac{dn_e}{dr^\prime}\right|^{-1},$ which for the exponential profile is simply ($1/\alpha$) and is independent of the normalization $n_0$.
Therefore, reasonable variations of the solar density model lead only to small quantitative shifts in the resonance position.
Numerical checks performed by varying the density normalization by up to $10\%$ produce changes that remain well below the quoted statistical uncertainties and do not alter the qualitative oscillation pattern or the main conclusions of the analysis.

\section{Results and Discussions}\label{sec5}
\begin{figure}[H]
	\centering
	\includegraphics[scale=0.60]{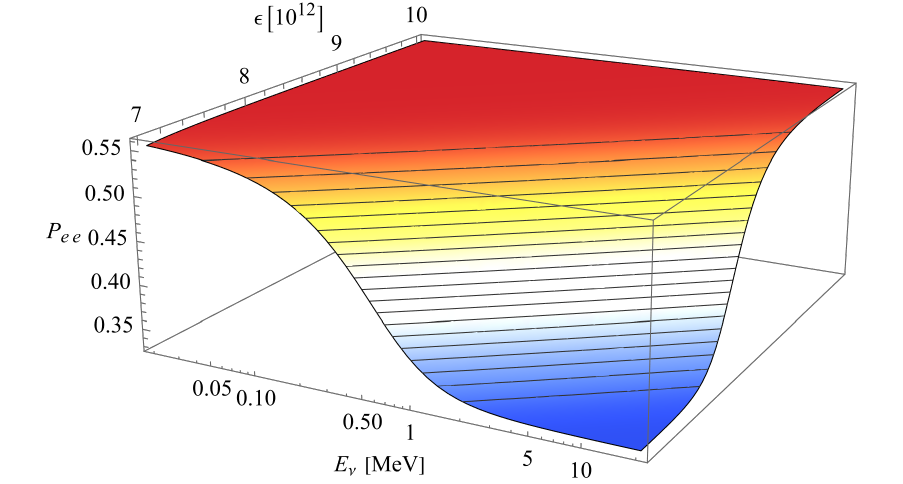}
	\caption{\footnotesize{This plot shows the 3D-dependence of the electron neutrino survival probability $P_{ee}$ on the coupling parameter $\epsilon$ and neutrino energy $E_\nu$, showing how non-standard interactions in teleparallel gravity affect solar neutrino oscillations.
	The pronounced drop in $P_{ee}$ for high-energy neutrinos highlights their potential for probing BSM physics via the LMA-MSW effect.‌}}
	\label{fig4}
\end{figure}

It is expected that non-standard interactions between neutrinos and a scalar field, as described by teleparallel gravity, can have a notable impact on the probabilities of neutrino oscillations in a vacuum. These interactions also modify the survival probability $P_{ee}(E_\nu)$ of solar electron neutrinos via the LMA-MSW effect as they travel through the Sun's matter.
Figure \ref{fig4} presents a 3D visualization of the survival probability $P_{ee}$ as a function of the coupling parameter $\epsilon$ and neutrino energy $E_\nu$, covering a range that includes neutrinos from $pp$, $pep$, $^7$Be, and $^8$B reactions.
At lower energies, such as those of $pp$ and $^7$Be neutrinos, $P_{ee}$ stays relatively high because the MSW effect has a weaker influence. However, for higher-energy $^8$B neutrinos, $P_{ee}$ drops significantly due to the enhanced flavor conversion caused by the MSW effect in regions of the Sun with high electron density.
The study of physics BSM in the context of solar neutrino flux is most effective for these high-energy $^8$B neutrinos, as demonstrated by the $P_{ee}$--$\epsilon$ relationship.

\begin{figure}[H]
	\centering
	\includegraphics[scale=0.45]{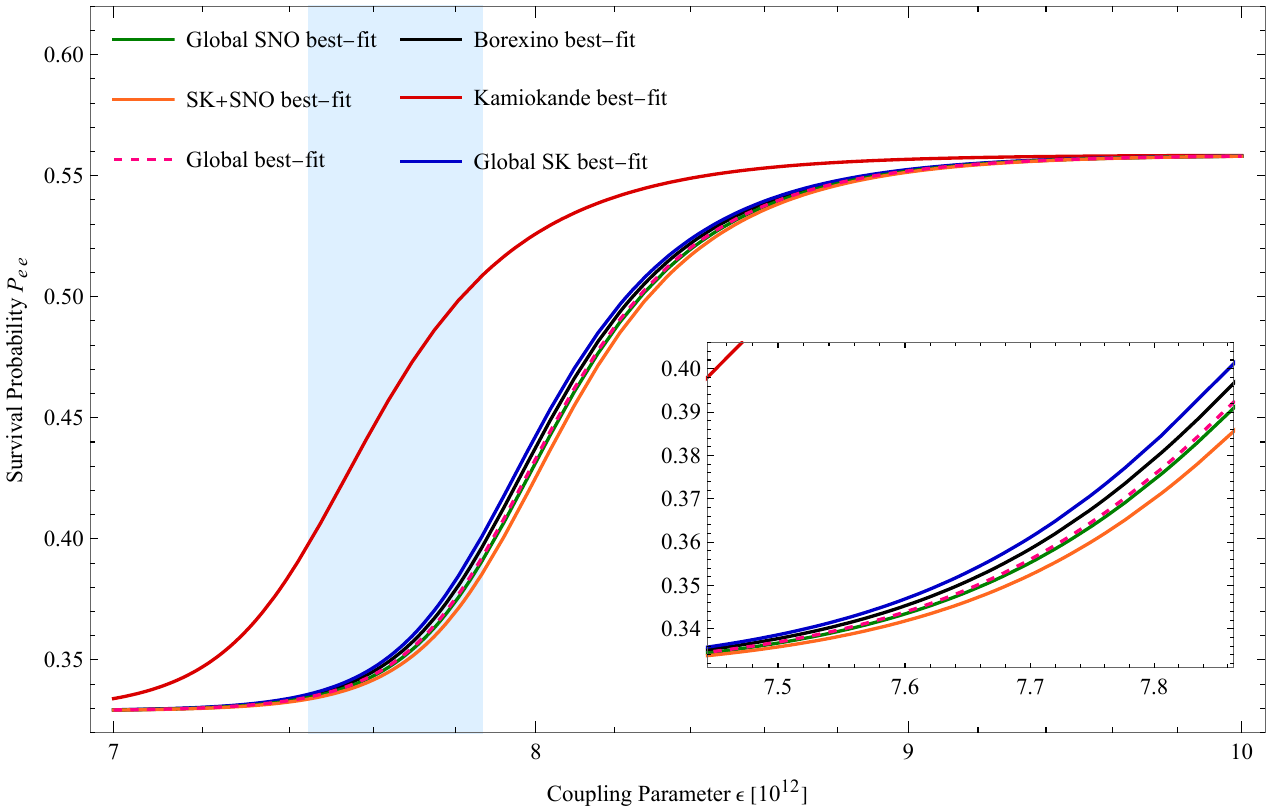}
	\caption{\footnotesize{Dependence of $P_{ee}$ on the $\phi$-$T$ coupling parameter $\epsilon$, using best-fit values from solar neutrino experiments (for $E_\nu \simeq 10 \text{MeV}$).
	The plot reveals that Kamiokande data is a potential outlier, deviating from other experimental results.}}
	\label{fig5}
\end{figure}

According to Fig.\ref{fig4}, figure \ref{fig5} is presented to illustrate the dependence of the solar neutrino survival probability on the scalar-torsion coupling parameter $\epsilon$.
The data employed in this figure correspond to the best-fit parameters derived from various solar neutrino experiments, as reported in Tables \ref{table1} and \ref{table2}.
The light-blue shaded region represents the range of best-fit values obtained from the aforementioned experiments, spanning from the minimum to the maximum value.
A notable observation from the inset is the close proximity of all curves, with the exception of the data associated with the Kamiokande case (the red curve).
Therefore, the Kamiokande data could be interpreted as outlier.

\begin{figure}[H]
	\centering
	\includegraphics[scale=0.50]{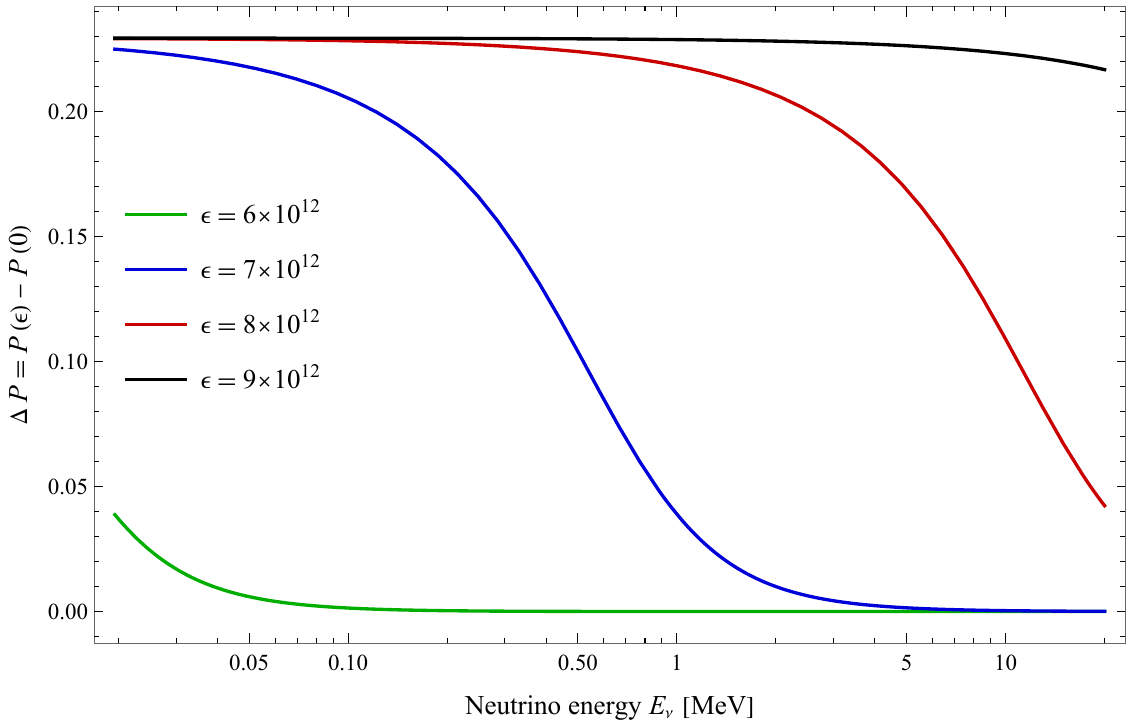}
	\caption{\footnotesize{This figure illustrates how a proposed neutrino-torsion non-standard coupling $\epsilon$ modifies the expected behavior of solar neutrinos as they propagate through the Sun and space, with the magnitude and nature of this modification being intricately linked to the neutrino energy $E_{\nu}$.}}
	\label{fig6}
\end{figure}

The difference in the survival probability of solar electron-neutrinos ($\Delta P = P(\epsilon) - P(0)$) compared to the standard scenario has been displayed in Figure \ref{fig6}, plotted against the neutrino energy $E_{\nu}$.
The standard scenario, where $\epsilon \to 0$, represents neutrino oscillations governed by the usual neutrino flavor conversion, i.e., vacuum oscillations and the MSW mechanism inside matter, but without this non-standard coupling to torsion.
The curve corresponding to $\epsilon \to 0$, which serves as our reference point, lies along the $E_\nu$-axis, indicating $\Delta P = 0$.
The other curves represent different, non-zero values of the coupling parameter $\epsilon \in \{6 \times 10^{12},7 \times 10^{12},8 \times 10^{12},9 \times 10^{12}\}$.
This plot clearly shows that as $\epsilon$ increases, the deviation $\Delta P$ from the standard probability becomes larger.
This means that stronger non-standard couplings have a more significant impact on how electron neutrinos survive their journey from the Sun to Earth.

Moreover, the effect of the coupling parameter $\epsilon$ is strongly dependent on the neutrino energy.
At lower energies (e.g., for $E_\nu \lesssim 1$ MeV), the deviation $\Delta P$ is relatively large for $\epsilon \gtrsim 7\times 10^{12}$.
This suggests that at these lower energies, the neutrino-torsion non-standard coupling causes a large increase in $P_{ee}$ compared to the standard model.
The modified MSW effect in solar matter is more dominant at these energies, allowing this subtle modification to be observed.
Although one might expect that the lowest-energy neutrinos—such as those from $pp$, $^7$Be, and $pep$ processes observed in the Borexino experiment—would serve as the most suitable candidates for detecting physics BSM, there is a critical challenge.
Experiments like Borexino, which are sensitive to low-energy neutrinos, also detect a range of other neutrino energies and are subject to contributions from various unavoidable background sources.
These include radioactive isotopes contaminating the scintillator material \cite{Agarwalla} as well as cosmic rays interacting in the Earth's atmosphere \cite{Kumaran}.
Such background effects can significantly influence the measured electron-neutrino survival probability, thereby potentially obscuring or distorting BSM signals.
Consequently, within the energy range accessible to these experiments, the exploration of BSM physics is most viable in the region dominated by the high-energy $^8$B solar neutrino flux.

Thus, for higher-energy neutrinos—such as those from the $^8$B process—the deviation $\Delta P$ attains its minimum value in each case.
This suggests that at these particular energy levels, the influence of non-standard couplings on the electron neutrino survival probability is reduced.
Notably, within the coupling strength range of $6 \times 10^{12} \lesssim \epsilon \lesssim 9 \times 10^{12}$, the magnitude of this deviation is more significant compared to other ranges, making it potentially detectable in future experimental studies.

\begin{figure}[H]
	\centering
	\includegraphics[scale=0.55]{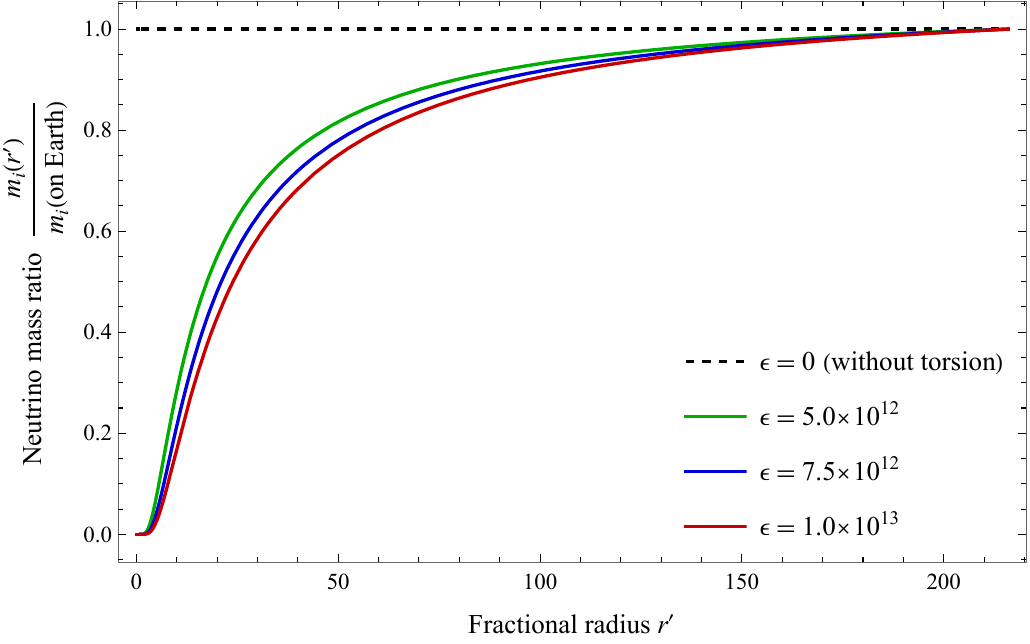}
	\caption{\footnotesize{This figure illustrates the ratio of neutrino mass $m_i(r^\prime)$ to terrestrial measurements, highlighting the variation in neutrino mass from the Sun's core to Earth under the influence of torsion-induced matter effects.
			The plot reveals how neutrino mass increases with distance and its dependence on the coupling parameter $\epsilon$, with larger values leading to a diminished increase in mass fraction outside the Sun.}}
	\label{fig7}
\end{figure}

Figure \ref{fig7} presents a plot depicting the ratio of neutrino mass $m_i(r^\prime)$ to the mass of neutrinos measured in terrestrial experiments, illustrating the variation in neutrino mass as they travel from the core of the Sun to the Earth. This variation is influenced by factors such as matter effects and potential new physics arising from non-minimal scalar field couplings to torsion, as described by the teleparallel gravity model.

Near the Sun's core, neutrino masses are exceedingly small, resulting in minimal flavor oscillations well within the solar medium. Beyond the Sun, neutrinos propagate through a vacuum, where flavor conversion is primarily determined by vacuum oscillations. The ratio of neutrino mass may demonstrate a significant increase as a function of distance, indicative of the interplay between neutrino mass and the scalar field. As the distance increases, the neutrino mass approaches a limiting standard value as $\epsilon \to 0$ (without torsion).

This figure effectively illustrates the dependence of neutrino mass on the non-minimal coupling parameter $\epsilon$. Curves corresponding to larger values of $\epsilon$ may exhibit a less pronounced increase in the fraction of neutrino mass outside the solar environment.

\begin{figure}[H]
	\begin{subfigure}{.5\textwidth}
		\centering
		\includegraphics[scale=0.39]{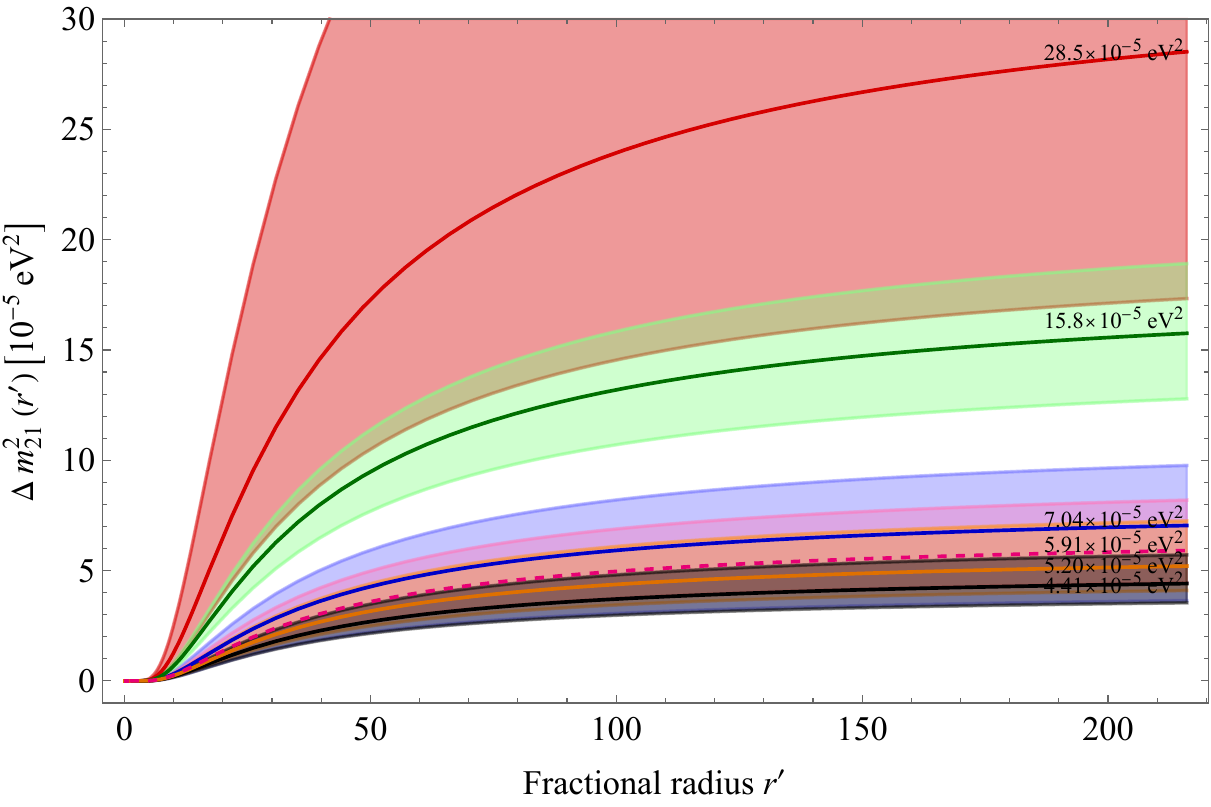}
		\label{fig5-a}
		\caption{\footnotesize{}}
	\end{subfigure}
	\begin{subfigure}{.5\textwidth}
		\centering
		\includegraphics[scale=0.39]{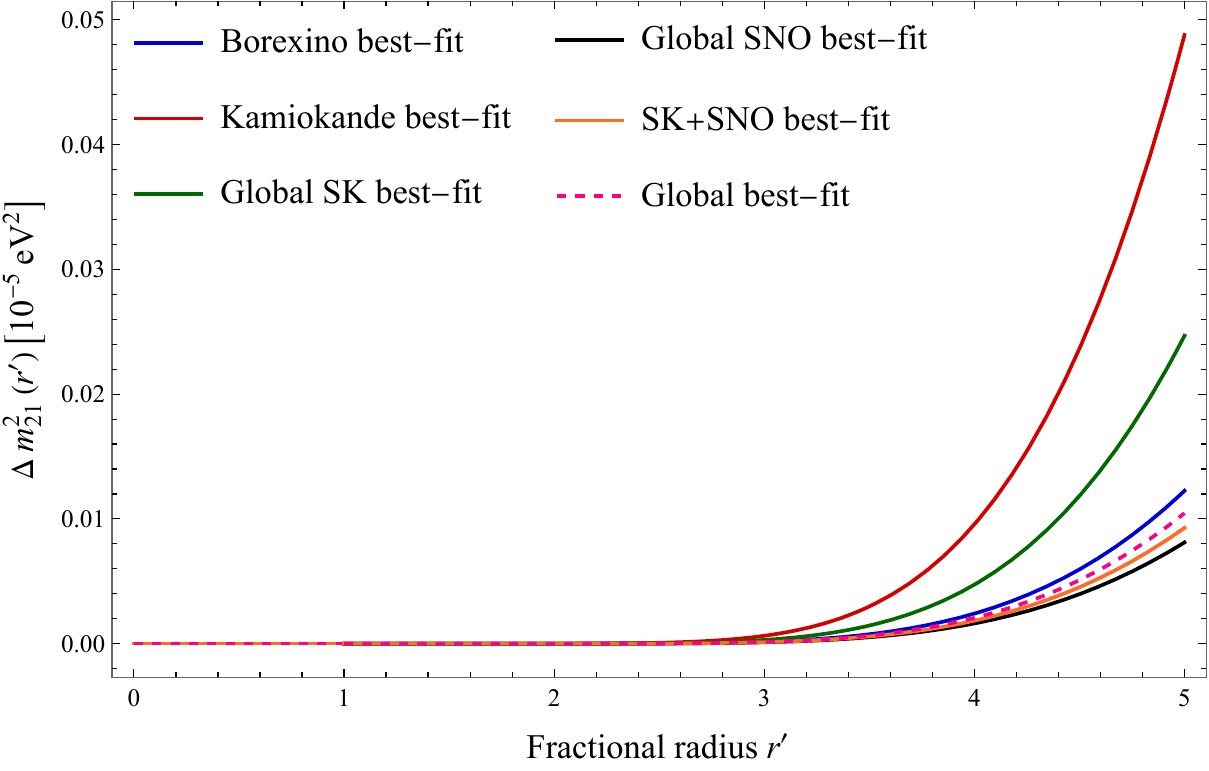}
		\label{fig5-b}
		\caption{\footnotesize{}}
	\end{subfigure}
	\caption{\footnotesize{Variation of the squared mass difference $\Delta m_{21}^2(r^\prime)$ as a function of dimensionless fractional radius ($r^\prime$).
			This would be a comparative analysis from Borexino, SNO, Kamiokande, SK, and global fits.
			The best-fit values of $\Delta m_{21}^{2}$ obtained from various analyses are as follows: Borexino yields $(7.04^{+2.72}_{-3.41}) \times 10^{-5} \,\text{eV}^2$; the global fit gives $(5.91^{+2.29}_{-1.45}) \times 10^{-5} \,\text{eV}^2$; the combined SK+SNO analysis provides $(5.20^{+2.05}_{-1.11}) \times 10^{-5} \,\text{eV}^2$; the global SNO analysis returns $(4.41^{+1.30}_{-0.87}) \times 10^{-5} \,\text{eV}^2$; and the global SK analysis yields $(18.9^{+3.16}_{-2.97}) \times 10^{-5} \,\text{eV}^2$.}}
	\label{fig8}
\end{figure}

The mass-squared difference $\Delta m^2_{21}$ between the neutrino mass eigenstates is a key parameter in the investigation of neutrino oscillations.
Understanding how $\Delta m^2_{21}$ changes with respect to the fractional radius helps provide a more detailed picture of neutrino behavior.
Figure \ref{fig8} shows the asymptotic $\Delta m_{21}^2$ values derived from each experiment and from the global fit, together with the corresponding $1\sigma$ uncertainty bands obtained by propagating the confidence intervals of $\epsilon$ and $\Delta\kappa_{21}^{\prime 2}$.
The uncertainty bands for most experiments overlap, indicating that the individual measurements are statistically consistent with one another and with the global fit.
The global fit yields $\Delta m_{21}^{2}=(5.91^{+2.29}_{-1.45})\times10^{-5}\ {\text{eV}}^{2}$, which is compatible within its uncertainties with the standard LMA-MSW value, $\Delta m_{21}^{2}\simeq7.5\times10^{-5}\ {\text{eV}}^{2}$ \cite{KamLAND,deSalas1,Esteban,deSalas2,Super-Kamiokande,Aharmim1,Bellini,Aharmim2}.
Among the individual analyses, Borexino yields $(7.04^{+2.72}_{-3.41})\times10^{-5}\ {\text{eV}}^{2}$, the combined SK+SNO analysis gives $(5.20^{+2.05}_{-1.11})\times10^{-5}\ {\text{eV}}^{2}$, and the global SNO fit gives $(4.41^{+1.30}_{-0.87})\times10^{-5}\ {\text{eV}}^{2}$.
Although their central values differ, the corresponding uncertainty intervals overlap and remain compatible with the LMA-MSW region.
These uncertainty intervals demonstrate that the apparent differences among the central values are not statistically significant once the experimental errors are taken into account.
The Kamiokande result, however, favors a larger central value of $\Delta m_{21}^{2}$ and exhibits the broadest allowed interval among all datasets, reflecting the limited statistics and larger experimental uncertainty of that pioneering experiment.
Overall, once the experimental uncertainties are taken into account, the inferred values of $\Delta m_{21}^{2}$ are mutually consistent for nearly all datasets, and the global result obtained within the torsion-modified framework remains compatible with the standard neutrino oscillation picture.
Reactor experiments such as KamLAND \cite{KamLAND} and JUNO \cite{JUNO} probe the same oscillation parameters, $\Delta m_{21}^{2}$ and $\theta_{12}$, as solar neutrino observations, but under terrestrial conditions.
In the present model, neutrino masses depend on the local scalar field value through the relation $m_i \propto \hat{\phi}^{2}$.
Since the scalar field approaches an approximately constant asymptotic value sufficiently far from the Sun, reactor experiments performed on Earth effectively probe the terrestrial quantity $\Delta m_{21}^{2}(r_\oplus) = \Delta\kappa_{21}^{\prime 2}\hat{\phi}^{4}(r_\oplus)$.
Using the best-fit parameters obtained from the global solar neutrino analysis, we find $\Delta m_{21}^{2} = (5.91^{+2.28}_{-1.46}) \times 10^{-5}\ {\text{eV}}^{2}$ (see also Fig. \ref{fig8}), which is compatible within uncertainties with the values inferred from KamLAND ($\Delta m_{21}^{2} = (7.53 \pm 0.18) \times 10^{-5} {\text{eV}}^{2}$) \cite{KamLAND} and JUNO ($\Delta m_{21}^{2} = (7.53 \pm 0.017) \times 10^{-5} {\text{eV}}^{2}$) \cite{JUNO-Data}.
Therefore, the present solar-neutrino constraints appear consistent with existing reactor-neutrino measurements and do not reveal any obvious tension with KamLAND and JUNO determinations of $\Delta m_{21}^{2}$.
Because KamLAND and especially JUNO provide highly precise values of $\Delta m_{21}^{2}$, their inclusion would primarily reduce the allowed range of $\Delta\kappa_{21}^{\prime 2}$ and, through its correlation with $\epsilon$, further tighten the constraints on the torsion-induced coupling parameter.
Such a global analysis is beyond the scope of the present work and is left for future investigation.

\section{Conclusion}\label{sec6}
This study investigates mass varying neutrino flavor conversion in a teleparallel gravity model where the neutrino mass term depends quadratically on a scalar field, producing a quartic scalar field term in the propagation equation.
Analytical and numerical solutions for a Schwarzschild spacetime show that the scalar field $\hat{\phi}$ increases radially outward to an asymptotic value (see section \ref{sec2}).
The torsion-scalar and scalar-neutrino couplings modify the MSW effect, influencing neutrino masses, mass-squared differences, and flavor transitions (see section \ref{sec3}).
Using solar neutrino data, the parameters $\epsilon$ and $\Delta \kappa^{\prime}_{21}$ are constrained, yielding results consistent with the LMA-MSW solution (see sections \ref{sec4} and \ref{sec5}).
While direct detection of model-specific effects is not yet possible, current gravitational experiments impose screening constraints, and future facilities like Hyper-Kamiokande \cite{HK1,HK2} and DUNE \cite{DUNE} can further test the model.
Overall, this work links neutrino physics with teleparallel modified gravity, demonstrating the utility of solar neutrino experiments in probing non-standard interactions.

\section*{Acknowledgement}
This work is based upon research funded by Iran National Science Foundation (INSF) under project No. 4036948.

\end{document}